\documentclass[useAMS,usenatbib]{mn2e}
\usepackage{graphicx}
\usepackage{color}
\usepackage{multirow}
\usepackage{amssymb,amsmath}
\usepackage{graphicx}
\usepackage{caption}
\usepackage{placeins}

\def\ps{$\rm km \,s^{-1}\,kpc^{-1}$}
\def\lv{\emph{l-v} }
\def\xy{\emph{x-y} }
\def\arcdeg{$^\circ$}
\def\kms{\,km\,s$^{-1}$}
\newcommand{\sech}{\mathrm{sech} \,}

\def\H2{$\rm H_2$}
\def\torus{\textsc{torus}}

\def\Robs{$R_{\rm obs}$}
\def\Vobs{$V_{\rm obs}$}
\def\lobs{$l_{\rm obs}$}

\title[The morphology of the Milky Way - I.]{The morphology of the Milky Way - I. Reconstructing CO maps from simulations in fixed potentials}

\author[A. R. Pettitt, C. L. Dobbs, D. M. Acreman and D. J. Price]{Alex R. Pettitt$^{1}$\thanks{E-mail:
alex@astro.ex.ac.uk}, Clare L. Dobbs$^{1}$, David M. Acreman$^{1}$ and Daniel J. Price$^{2}$\\
$^{1}$School of Physics and Astronomy, University of Exeter, Stocker Road, Exeter EX4 4QL, UK\\
$^{2}$Monash Centre for Astrophysics (MoCA), School of Mathematical Sciences, Monash University, VIC 3800, Australia\\
}
\begin{document}

\date{\today}

\pagerange{\pageref{firstpage}--\pageref{lastpage}} \pubyear{2013}

\maketitle

\label{firstpage}

\begin{abstract}
We present an investigation into the morphological features of the Milky Way. We use smoothed particle hydrodynamics (SPH) to simulate the interstellar medium (ISM) in the Milky Way under the effect of a number of different gravitational potentials representing spiral arms and bars, assuming the Milky Way is grand design in nature. The gas is subject to ISM cooling and chemistry, enabling us to track the evolution of molecular gas. We use a 3D radiative transfer code to simulate the emission from the SPH output, allowing for the construction of synthetic longitude-velocity (\emph{l-v}) emission maps as viewed from the Earth. By comparing these maps with the observed emission in CO from the Milky Way, we infer the arm/bar geometry that provides a best fit to our Galaxy. We find that it is possible to reproduce nearly all features of the \emph{l-v} diagram in CO emission. There is no model, however, that satisfactorily reproduces all of the features simultaneously. Models with 2 arms cannot reproduce all the observed arm features, while 4 armed models produce too bright local emission in the inner Galaxy. Our best-fitting models favour a bar pattern speed within 50-60\ps\, and an arm pattern speed of approximately 20\ps, with a bar orientation of approximately 45\arcdeg\,and arm pitch angle between 10\arcdeg-15\arcdeg.
\end{abstract}


\section{Introduction}
Despite decades of research, the structure of our own Galaxy still remains a mystery. Whilst we are able to discern a wealth of information regarding the structure of galaxies in the night sky (e.g. \citealt{1959HDP....53..275D}, \citealt{2007AJ....134..579F}, \citealt{2011ApJ...737...32E}, \citealt{2013MNRAS.435.2835W} etc.), the simplest morphological questions about our Galaxy are still in dispute (e.g. \citealt{2004ApJS..154..553S,2005AJ....130..569V,2008ASPC..387..375B} and \citealt{2012MNRAS.422.1283F}). Our location in the Galactic disc means that it is difficult to discern the number and shape of the spiral arms, and the structure of the inner bar.

The earliest works mapping the Milky Way utilised the detection of the HI 21-cm line (e.g. \citealt{1958MNRAS.118..379O} and \citealt{1962MNRAS.123..327K}) and gave tantalising early evidence of the Milky Way's spiral structure. However, these maps require some underlying model of Galactic rotation, and are incapable of reliable detection of emission beyond the Galactic centre. Ionised hydrogen and OB stars are also commonly used to map out Galactic structure, e.g. \citet{1976A&A....49...57G}, \citet{1987A&A...171..261C} and \citet{2003ApJ...582..756K}. Determining accurate  distances to these sources requires breaking the ``kinematic distance ambiguity", which makes distance determinations in the inner Galaxy problematic (see \citealt{2009ApJ...699.1153R} for a discussion). 

\begin{figure}
\includegraphics[width=84mm]{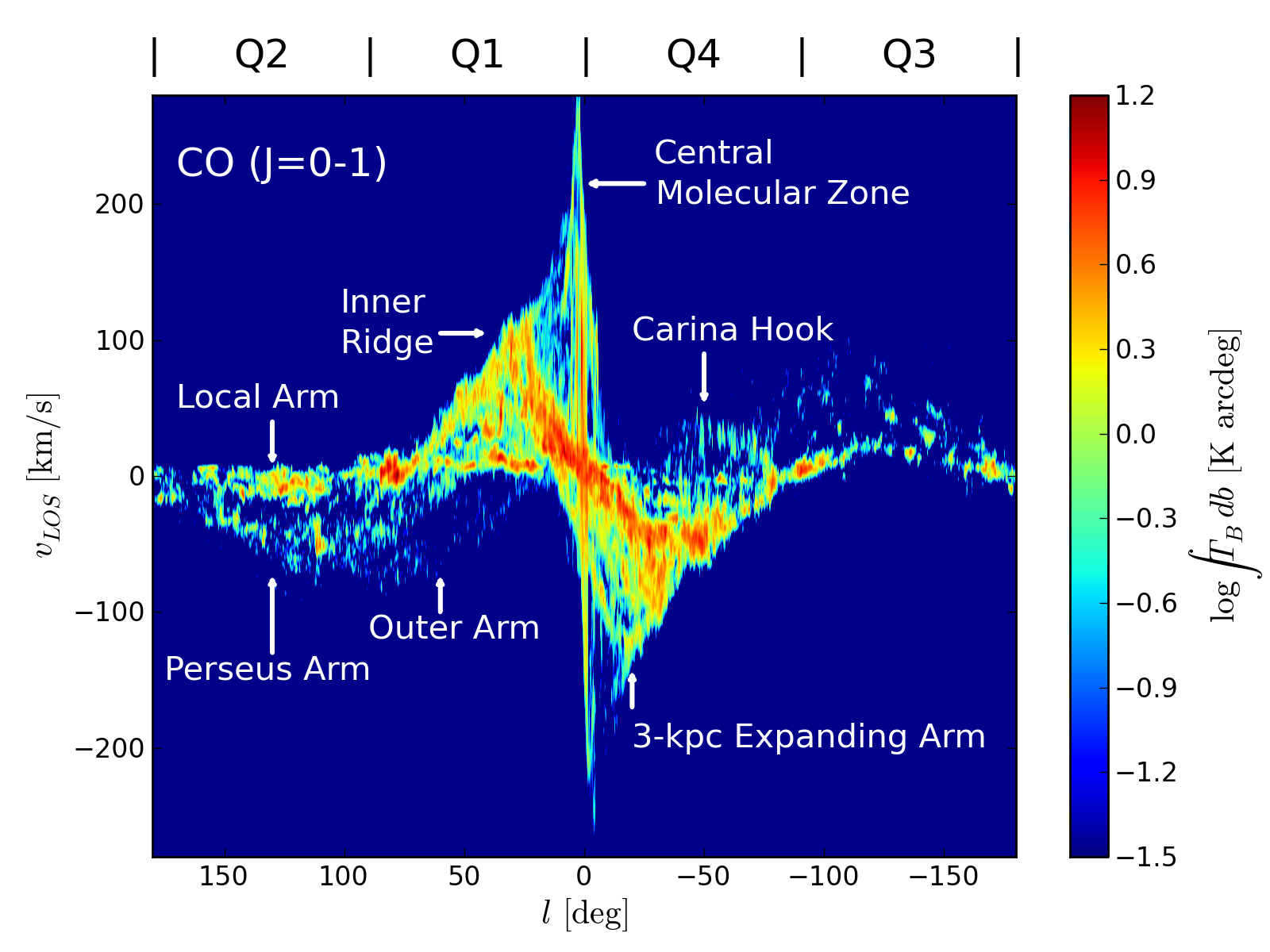}
 \caption{Longitude-velocity map of brightness temperature of the CO (J=0-1) transition (\citealt{2001ApJ...547..792D}), with major arm features labelled. We integrate the CO emission over $\pm2^\circ$ latitude in order to show weaker features. Q1-4 indicates the position of Galactic quadrants.}
\label{DameLV}
\end{figure}

There is not yet a consensus on a single Galactic spiral model \citep{1985IAUS..106..255E,1985IAUS..106..283L}. Some studies favoured a 4-armed spiral structure (e.g. \citealt{1976A&A....49...57G}, \citealt{1993ApJ...411..674T}), some a 2-armed structure (e.g. \citealt{1970IAUS...38..126W}), and some a ring of material in the inner Galaxy (e.g. \citealt{1977ApJ...217L.155C}). 
Work in recent years has still been unable to converge on a preferred model. A study of star forming complexes throughout the Galactic disc by \citet{2003A&A...397..133R} displayed a preference for a 4, rather than 2 or 3 armed model. 
However, their best fit still has a large amount of scatter, especially 3\,kpc inside of the solar position and behind the Galactic centre. 
Maps of the inner and outer Galaxy by \citet{2009A&A...499..473H} and \cite{2006Sci...312.1773L} find a 2 armed spiral insufficient to fit their data, favouring 3 and 4 armed models. The best-fitting spirals to the map of \citet{2009A&A...499..473H} give a fairly tightly wound pattern with a pitch angle of around 10\arcdeg, while that of \cite{2006Sci...312.1773L} favours a much looser spiral with a pitch of around 20\arcdeg. Furthermore, the \citet{2009A&A...499..473H} map is best fit by an asymmetric spiral model of fixed pitch angle, while that of \cite{2006Sci...312.1773L} is best fit by a polynomial arm model up to fourth order. This evidence could lead one to think the structure may differ significantly between the outer and inner Galaxy \citep{2011MSAIS..18..199E}. The maps of \cite{2006Sci...312.1773L} and \citet{2009A&A...499..473H} were reanalysed along with 2MASS star frequencies by \citet{2012MNRAS.422.1283F}, who conversely found that a 2-armed spiral with a very tight pitch of about 5\arcdeg\,provided a good fit to all the data. A separate synthesis of data by \citet{2011ARep...55..108E} finds a 4-armed structure most plausible in the inner Galaxy.
A statistical analysis of several other studies in the literature is performed in \citet{2002ApJ...566..261V,2005AJ....130..569V,2008AJ....135.1301V}, where the author favours a symmetric 4-armed spiral model.

Recent analysis of red-clump giant star counts from the \emph{Spitzer} GLIMPSE survey by \citet{2005ApJ...630L.149B} and \citet{2009PASP..121..213C} give a distance-independent view of Galactic structure by measuring source counts as a function of longitude. Higher concentrations of stars are believed to indicate the presence of spiral arm tangents and bar structures. The asymmetry in longitude of the GLIMPSE data towards the Galactic centre is an indicator of a bar like structure. The authors attribute this to a 4\,kpc long, thin bar orientated at $\theta_b\approx 45^\circ$, also seen by \citet{2000MNRAS.317L..45H}. This is at odds with the normal 3\,kpc long $\theta_b\approx 20^\circ$, so called \emph{COBE} DIRBE bar believed to reside in the Galactic centre (\citealt{1991ApJ...379..631B}, \citealt{1994ApJ...425L..81W}, \citealt{1997MNRAS.288..365B}, \citealt{2002ASPC..273...73G} and references therein). The DIRBE bar is believed to be a more classical bulge-like structure in the central Galaxy with a triaxial peanut/boxy density distribution. The spiral tangents as seen by GLIMPSE clearly show the Scutum, near-3\,kpc and Centaurus arms, but there is little evidence of the Norma and Sagittarius features \citep{2009PASP..121..213C}. 

A plausible reasoning behind the discrepancy in different arm numbers is that different sources trace different structures. Some studies indicate that the old stellar population ($J$ and $K$ band emission) is best fit by a 2-armed spiral, while the gas/dust emission is best fit by a 4-armed spiral. This is seen from maps of emission/counts as function of longitude seen in the \emph{COBE} DIRBE data (\citealt{2000A&A...358L..13D}, \citealt{2001ApJ...556..181D}) and \emph{COBE} FIRAS NII/CII data \citep{2010ApJ...722.1460S}. It is possible that the observed 4-armed spiral in the gas is being driven by a 2-armed spiral in the old stellar population \citep{2004MNRAS.350L..47M}. Similar tangent maps by \citet{2012ApJ...747...43B} show cold dust emission (870$\mu$m) that peaks in the expected position of tangencies of a 4-armed spiral, seemingly at odds with the flat longitude-count maps of young stellar objects seen by \emph{Spitzer}.

Many different individual studies have also produced longitude-velocity (\emph{l-v}) maps of our Galaxy in a number of ISM tracers. These include HI (\citealt{1970IAUS...38..397B}, \citealt{1982ApJ...259L..63K}, \citealt{2005A&A...440..775K}, \citealt{2007AJ....134.2252S}, \citealt{2012ApJS..199...12M}), HII (\citealt{1987A&A...171..261C}), CO (\citealt{1980ApJ...239L..53C}, \citealt{1986ApJ...305..892D}, \citealt{2001ApJ...547..792D}, \citealt{2006ApJS..163..145J}) and CII (\citealt{2013A&A...554A.103P}). An \lv map of the full Galactic plane in molecular emission (specifically the CO (J=0-1) transitions) is shown in Figure \ref{DameLV}, created from the data presented in \citet{2001ApJ...547..792D}. Strong regions of emission in these \lv maps are believed to trace out Galactic spiral structure, due to the associated higher stellar and gaseous densities. The map in Fig.\,\ref{DameLV} clearly shows some well known arm structures, such as the Perseus and Carina arms. Two features that appear in molecular emission, but are absent in atomic emission, are the bright inner ridge ($l$=40\arcdeg\,to -40\arcdeg\,, $|v_{los}|<$ 120\kms, mainly the Scutum-Centaurus-Crux, SCC, arm) and the central molecular zone (CMZ, $|l|<$10\arcdeg\,, $|v_{los}|<$ -280\kms).

There have been numerous numerical studies focusing on fitting models to \lv data. These have focused on modelling both the inner and outer Galaxy using simulations including bar and spiral features (e.g. \citealt{1999MNRAS.304..512E}, \citealt{1999A&A...345..787F}, \citealt{2004ApJ...615..758G}, \citealt{refId0}, \citealt{2010PASJ...62.1413B}, \citealt{2013MNRAS.428.2311K}). Studies have also examined the so-called Galactic molecular ring (\citealt{2011MNRAS.418.2508M}, \citealt{2012MNRAS.421.2940D}). 
However, none have yet attempted to create actual emission maps of the entire Galactic plane, instead relying on simply projecting arm and bar features from \xy to \lv space.

In previous work synthetic Galactic plane observations have been created of a single quadrant of a simulated spiral galaxy in HI (\citealt{2010MNRAS.407..405D}, \citealt{2012MNRAS.422..241A}), finding reasonable agreement between the strength and location of emission. The study presented here is an attempt to create synthetic emission maps of the ISM (principally in CO) for the entire Galactic disc, with the aim of discerning the spiral/bar structure of our Galaxy by comparison to the observational data shown in Fig.\,\ref{DameLV}. This work is in a similar vein to \citet{2012MNRAS.421.2940D}, except we use numerical simulations to produce realistic ISM structures, subject to some spiral/bar potential, rather than assuming the gas directly traces the locations of the perturbations. This allows for the capture of shocks produced during the passages through spiral arms (\citealt{1969ApJ...158..123R}, \citealt{2006MNRAS.371.1663D}), and the tracking of important properties of the gas without assuming some global distribution model (e.g. the molecular content or density distribution).

This paper is organised as follows. In Section \ref{Numerics} we discuss the various numerical simulations and the methodology behind the construction of the \lv maps. In Section \ref{Results} we discuss the results of the simulations and the resulting \lv maps. This is split into separate subsections describing; general features of the CO maps, bar, arm and bar+arm models. The comparison between these various models and their implications are discussed in Section \ref{Discussion}, with concluding remarks in Section \ref{Conclusion}.

\section[]{Numerical simulations}
\label{Numerics}
We use smoothed particle hydrodynamics (SPH) to simulate the flow of ISM gas in the Milky Way. SPH is a Lagrangian fluid formulation where each fluid packet, or particle, has a density that is smoothed over the neighbouring particles by a smoothing length, $h$, which is related to density by
\begin{equation}
\rho = m \left( \frac{\eta}{h} \right)^3
\end{equation}
in 3D, where $m$ the mass of the particles and $\eta$ is some constant chosen to set the mean number of neighbours for each particle (e.g. \citealt{2012JCoPh.231..759P}). We use fixed analytic potentials to represent the stellar mass distribution using the SPH code \textsc{phantom} \citep{2010MNRAS.405.1212L,2010MNRAS.406.1659P}. \textsc{phantom} is a low-memory, highly efficient SPH code written especially for studying non-self-gravitating problems. Particles in \textsc{phantom} have individual smoothing lengths and time steps, and are integrated using a leapfrog algorithm.

\begin{figure}
 \includegraphics[trim = 0mm 0mm 0mm 0mm,width=84mm]{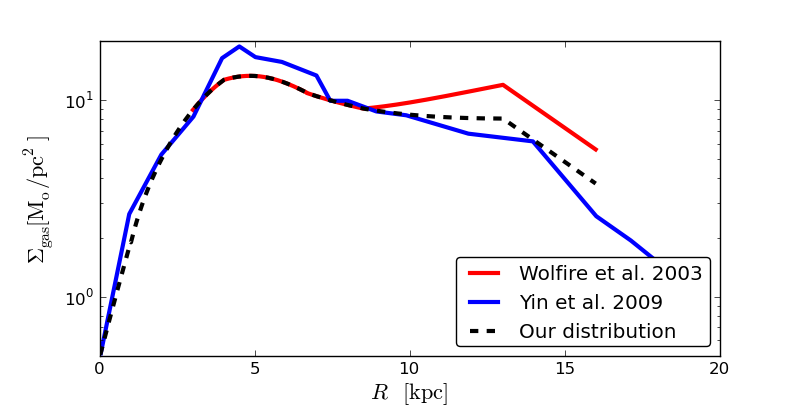}
  \caption{Initial surface density profile used in our simulations, shown as the black dashed line. The red line is the model of \citet{2003ApJ...587..278W} and the blue line is the data from \citet{2009A&A...505..497Y}, taken from \citet{1999MNRAS.307..857B}, which extends to the Galactic centre. Note that we only set gas out to a radius of 13\,kpc.}
  \label{SurfDen}
\end{figure}

For the simulations shown here we do not include stellar feedback, self-gravity or magnetic fields. Our primary aim is to investigate a large parameter space of possible potentials. The stellar gravitational field will be the primary driving force in the global distribution of ISM gas and we leave additional physical processes to a future study. The ISM gas is initially distributed in the Galactic plane, with a disc height of 0.4 kpc. The initial vertical distribution is of little importance, as all the gas falls into a disc of height 0.1 kpc after only 50Myr of evolution.
The initial surface density profile is chosen to match observational data. This is based on the functional form of \citet{2003ApJ...587..278W}. We impose a flat distribution instead of the authors slightly increasing density profile in the 8.5-13.0\,kpc region so that our surface density is not increasing near the edge of the disc. Some observations suggest that the distribution is effectively flat from 5 to 15\,kpc (see \citealt{doi:10.1146/annurev-astro-082708-101823} and references therein). We also extrapolate the density profile to the Galactic centre, using the data from \citet{2009A&A...505..497Y}. Our initial surface density profile is shown in Fig.\,\ref{SurfDen}, and gas in our simulations is set placed between 0 and 13\,kpc. Our choice of 13 kpc ensures the major spiral features recorded in the literature are included as far out as the Outer and Perseus arms. Though there is some evidence for weak spiral structure extending to 20 kpc \citep{2006Sci...312.1773L}, this would have required a large increase in particle number to achieve the same resolution, whilst any features at such large radius would have little influence on our results.
Integration of this surface mass distribution gives a total gas mass of approximately $8\times 10^{9}\, \rm M_\odot$, corresponding to an average ISM density of approximately $1\rm \, g\,cm^{-3}$ or 15$\rm M_\odot\, pc^{-2}$. Our fiducial resolution is 5 million SPH particles. A short resolution study is presented in Appendix \ref{ResStudy}.

\subsection{Chemistry and Cooling}
The gas in our simulations evolves according to an adiabatic equation of state and is subjected to ISM heating and cooling. We include ISM heating and cooling adapted from \citet{2007ApJS..169..239G} which includes the effects of cooling from atomic lines, photoelectric heating, fine-structure cooling and heating from cosmic rays.

Each SPH particle has a chemical abundance array that is updated along with the various hydrodynamical properties, allowing for the evolution of the Galactic atomic and molecular content. All particles are initially composed of 100\% HI. The formulation of \H2 chemistry is taken from \citet{2004ApJ...612..921B}, the implementation of which is described in \citet{2008MNRAS.389.1097D}. The \H2 is formed on the surface of dust grains, and is destroyed by photodissociation (a function of visual extinction and \H2 column density) and cosmic rays. In addition to \citet{2008MNRAS.389.1097D} we also follow the evolution of CO to enable the construction of synthetic molecular emission maps. We use the CO rate equations of \citet{1997ApJ...482..796N}. This treatment ignores any tracking of intermediate species and simply evolves CII to CO via formation of a hydrocarbon intermediate step. The hydrocarbon is created by interaction with $\rm H_2$ via;
\begin{equation}
\rm CII + H_2 \rightarrow CH_2^+ +\gamma,
\end{equation}
at a rate of $k_0$. This is assumed to be the slowest step in the process of forming CO. The CH$_2^+$ then converts to CH and CH$_2$ (denoted collectively as CH$_{\rm X}$) on very short time-scales using H$_2$. The resulting hydrocarbon is then allowed to react with OI to create CO;
\begin{equation}
\rm OI + CH_X \rightarrow CO + H_X + \gamma,
\end{equation}
at a rate of $k_1$. The rate equation encompassing these processes and the evolution of CO number density, $n_{CO}$, used by \citet{1997ApJ...482..796N} is 
\begin{equation}
\dot{n}_{CO}=k_0 n_{H_2}n_{CII}\beta - \zeta_{CO}n_{CO}.
\label{COeq}
\end{equation}
The $\beta$ factor dictates how much CH$_{\rm X}$ successfully transforms into CO and is given by
\begin{equation}
\beta=\frac{k_1 n_{\rm OI}}{k_1 n_{\rm OI}+\zeta_{\rm CH_X }}.
\label{betaEq}
\end{equation}
Where $n_X$ is the number density of species $X$ and $n$ is the total number density. The CO and $\rm CH_X$ is depleted through photodissociation at rates of $\zeta_{\rm CO}$ and $\zeta_{\rm CH_X}$ which depend on the species column densities, $N_X$, and visual extinction. The fine details and numerical constants for these reactions can be found in \citet{1997ApJ...482..796N} and \citet{2012MNRAS.421..116G}. We calculate column densities using the same method as \citet{2008MNRAS.389.1097D} where we approximate $N_X=n_X l_{ph} $ using a typical distance to a B0 star of $l_{ph}=30\rm\,pc$. We simply evolve CII and OI abundance as either being in their original state or as CO. We maintain the initial CII and OI abundances for the various cooling processes.

In previous work \H2 chemistry is subcycled inside main hydrodynamic timesteps due to the much shorter evolutionary time-scale \citep{2008MNRAS.389.1097D}. Here the CO formation is also subcycled along with the \H2, as it is directly coupled via Eq. \ref{COeq}. We also allow the CO formation to subcycle inside the \H2/HI chemical subcycle if required. 

An in depth study of the many other alternative models of ISM CO formation in the literature was performed by \citet{2012MNRAS.421..116G}. They find that the approach of \citet{1997ApJ...482..796N}, while simplistic compared to others, is good enough for tracing the global CO distribution in large scale simulations.

\subsection{Galactic potentials}
\subsubsection{Axisymmetric potential}
We use a combined bulge-halo-disc potential, $\Phi_{dbh}$, to reproduce the observed rotation curve, rather than assuming some simplified flat profile. Our axisymmetric rotation curve is shown in Fig.\,\ref{MWRC}  and described in Appendix \ref{AppAxi}. The velocities of the SPH particles are initialised from $\Phi_{dbh}$ with some additional small scale dispersion of $5$\kms.

\begin{figure}
 \includegraphics[trim = 0mm 0mm 0mm 0mm,width=84mm]{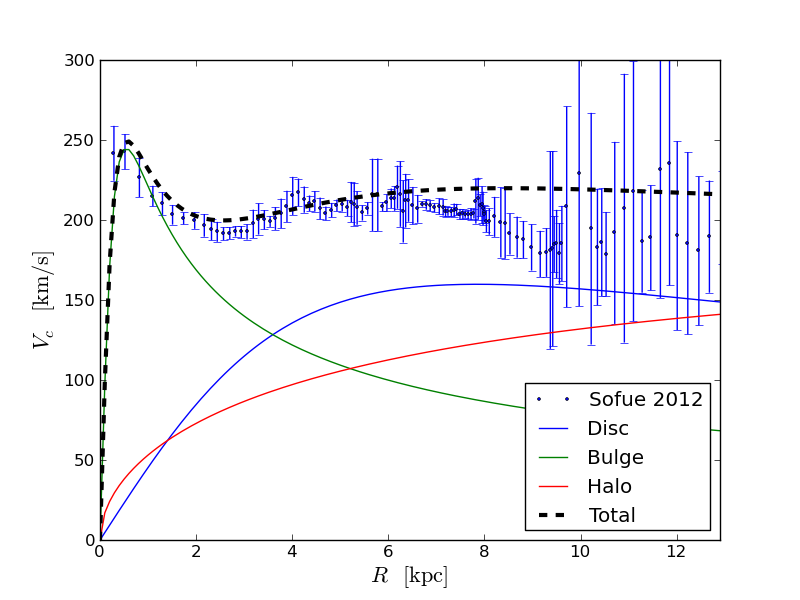}
  \caption{Rotation curve used in our simulations resulting from axisymmetric galactic potentials with observed rotation curve data from \citet{2012PASJ...64...75S}. The dashed line is the combined bulge-disc-halo model from \citet{1991RMxAA..22..255A} shown individually in green-blue-red respectively (see Appendix \ref{AppAxi} for details).}
  \label{MWRC}
\end{figure}

\begin{figure}
 \includegraphics[trim = 0mm 0mm 0mm 0mm,width=84mm]{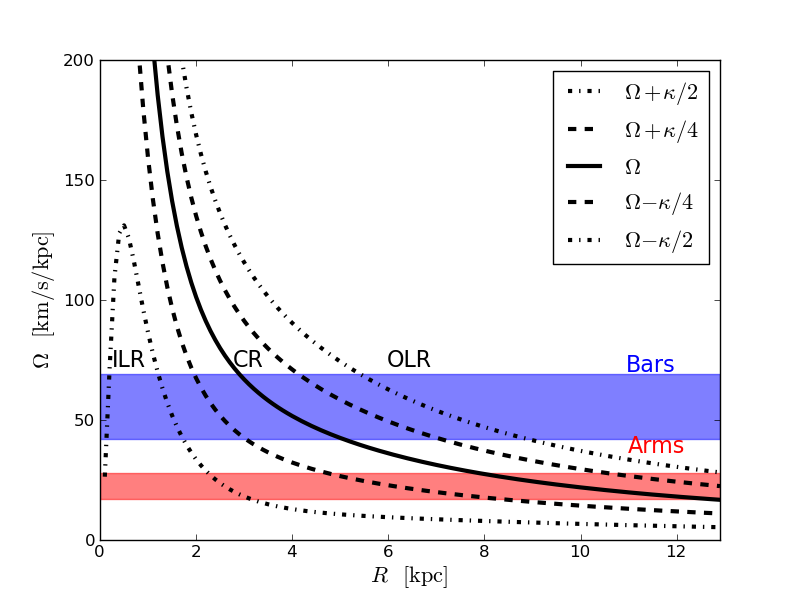}
  \caption{Rotation speeds from our adopted Milky Way rotation curve shown in Fig.\,\ref{MWRC}. The dashed and dot-dashed lines show the 4:1 and 2:1 resonances calculated from the epicycle frequency, $\kappa$. Upper and lower shaded regions show the possible region encompassed by the arm and bar pattern speeds, with maxima and minima from \citet{2011MSAIS..18..185G}.}
  \label{MWLR}
\end{figure}

The spiral and bar features are produced by subjecting the gas to further stellar potentials. When using fixed analytic potentials the structure of the Milky Way is assumed to be that of a grand design galaxy, driven by some stable stellar density wave. The potentials used here have a constant strength throughout the simulation. The radial extent of structures is determined by the location of the inner and outer Lindblad resonances, ILR and OLR, which are in turn determined by the pattern speed of the density wave, $\Omega_b$ or $\Omega_{sp}$. The frequencies resulting from our rotation curve in Fig.\,\ref{MWRC} are shown in Fig.\,\ref{MWLR}. For example, a 4-armed spiral perturbation with a pattern speed of 20\,\ps\, has ILR, OLR and CR (co-rotation radius) located at a radius 7.0, 14.4 and 10.9\,kpc respectively, shown by where the $\Omega \pm \kappa/4$ and $\Omega$ lines cross 20\ps\,in Fig.\,\ref{MWLR}. 

Bar and spiral potential parameters we choose to vary are summarised in Table \ref{POTparams}. These include the pitch angle of spiral arms ($\alpha$), the number of spiral arms ($N$), and the pattern speed of the bar and arms ($\Omega_{b},\Omega_{sp}$). We also investigate the effects on altering the strength of the potential perturbations, though we only use two separate values for the arm and bar components. The orientation of the bar/arm features to the observer ($l_{\rm obs}$), and the observer's velocity and Galactocentric distance ($V_{\rm obs}, R_{\rm obs}$) are also investigated but these will be varied during the construction of \lv maps. Our choice of parameters is broad and numerous to allow for an unbiased study, with as little recourse as possible to previous findings. There is both observational and numerical evidence for different pattern speeds for the arm and bar components in our Galaxy (e.g. \citealt{2011MSAIS..18..185G} and \citealt{1988MNRAS.231P..25S}).

\begin{table}
 \caption{Variable parameters of the simulations, including those of the arm/bar potentials and those used in defining the observer coordinates. Parameters in bold define the refined parameter space used in calculations with both bar and arm potentials.}
 \begin{tabular}{@{}lcc}
  \hline
  Term & Description & Values \\
  \hline
  \hline
  $\Omega_{b}$ & Bar pattern speed & 20, 40, \textbf{50}, \textbf{60}, 70\ps \\
  $\theta_{b}$ & Bar orientation & 0\arcdeg, 10\arcdeg, \ldots, 50\arcdeg, 60\arcdeg \\
  $\Omega_{sp}$ & Arm pattern speed & 10, \textbf{15}, \textbf{20}, 25, 30\ps \\
  $\alpha$ & Arm pitch angle & 5\arcdeg, \textbf{10\arcdeg}, \textbf{12.5\arcdeg}, \textbf{15\arcdeg}, 20\arcdeg\\
  $N$ & Number of arms   & \textbf{2}, \textbf{4} \\
  $|\Phi_{sp}|$ &Relative arm& {$\times$}\textbf{1}, $\times 2$ \\
  &potential strength& \\
  \hline
  $R_{\rm obs}$ & Radial position& 7, 7.5, 8, 8.5, 9 kpc \\
   & of the observer & \\
  $V_{\rm obs}$ & Circular velocity& 200, 205, \ldots, 225, 230 km\,s$^{-1}$ \\
   & of the observer & \\
  $l_{\rm obs}$ & Azimuthal position& 0\arcdeg, 10\arcdeg, \ldots, 350\arcdeg, 360\arcdeg \\
 &of the observer& \\
    \hline
 \end{tabular}
 \label{POTparams}
\end{table}

\subsubsection{Bars}
We employ two separate bar potentials to see which functional representation best matches the \lv features of our Galaxy. The first is a commonly used sinusoidal perturbation of the Galactic disc. We employ the specific form of \citet{2001PASJ...53.1163W};
\begin{equation}
\Phi_{r,W}(r,\phi) = \Phi_0 \cos \left( 2\left[ \phi+\Omega_b t \right] \right) \frac{\left(r/r_c\right)^2}{\left( \left(r/r_c\right)^2 + 1\right)^2},
\end{equation}
where $\Phi_0=\epsilon V_{0}^2 \sqrt{27/4}$, $\epsilon=0.05$ and $V_0=220$\kms. We employ two different values of the bar core radius, $r_c$, either 2\,kpc or $\sqrt{2}$\,kpc (used in \citealt{1994ApJ...437L.123W} and \citealt{2001PASJ...53.1163W} respectively). We will refer to these as the WK and WKr2 bars respectively throughout this paper. A measurement of the inner drop off radius of the bar potential, $r_c$ here also determines the strength of the potential, and so these values enable us to investigate the effect of the strength of the bar.

Another bar we employ is that of \citet{1992ApJ...397...44L}, referred to hereafter as the LM bar. The authors provide a bar model that is simply a softened line of gravitational potential. While not physically a bar, i.e. not the result of some density profile, the effect on the gas is still that of a non-axisymmetric perturbation. This potential is aligned with the x-axis by definition so we apply coordinate transforms to the positions and accelerations to simulate the rotation of the bar. The potential is given by
\begin{equation}
\Phi_r(x,y,z)=\frac{GM_r}{2a}\ln\left(  \frac{x-a+T_-}{x+a+T_+} \right),
\end{equation}
where $T_{\pm}=[(a\pm x)^2 + y^2 +(b+\sqrt{c^2+z^2})^2) ]^{1/2}$, with $a,b$ and $c$ roughly corresponding to the bar semimajor and minor axes respectively. We adopt a bar mass of $6.25\times 10^{10}\rm M_\odot$ as used by \citet{1999ApJ...513..242L} for the same potential.

We also initially included the bar model of \citet{2012MNRAS.427.1429W} which incorporates a `boxy' or `peanut' shaped bulge/bar. However, upon incorporating the potential into our simulations it became clear it showed very little difference compared to those models already discussed above. This is likely due to the boxy nature of the bar being predominantly in the vertical plane, and our simulations are effectively only considering minor motions in the vertical direction.

\subsubsection{Arms}
In \citet{2006MNRAS.371.1663D} a logarithmic spiral potential from \citet{2002ApJS..142..261C} was used, hereafter referred to as CG arms. This potential takes the form
\begin{equation}
\begin{aligned}
 &\Phi_{sp}(r,\phi,z)= 4\pi Gh_z\rho_o \exp{\left(-\frac{r-r_o}{R_s}\right)} \sum_n^3\bigg\{ \frac{C_n}{K_n D_n}\bigg.\\ 
 &\bigg. \times  \left[ \sech \left(\frac{K_n z}{\beta_n}\right)  \right]^{\beta_n}   \cos\left( N\left[\phi -t\Omega_{sp}-\frac{\ln(r/r_o)}{\tan(\alpha)}\right]\right) \bigg\}
\end{aligned}
\end{equation}
where 
\begin{equation}
K_n=nN/r\sin(\alpha),
\end{equation}
\begin{equation}
D_n = \frac{1+K_nh_z+0.3(K_n h_z)^2}{1+0.3K_n h_z},
\end{equation}
\begin{equation}
\beta_n=K_nh_z(1+0.4K_n h_z),
\end{equation}
and the constants are the same as those used in \citet{2006MNRAS.371.1663D}, namely $h_z=0.18$kpc, $R_s=7$kpc, $r_0=8$kpc, $C=(8/3\pi,1/2,8/15\pi)$ and a fiducial spiral density of $\rho_0=$1\,atom\,cm$^{-3}$.
These spiral arms take the form of a three part sinusoidal perturbation that exponentially decay with increasing radius.

We also implement the logarithmic spiral perturbation of \citet{2003ApJ...582..230P} due to its apparent effectiveness at creating four armed spiral patterns in the ISM gas from only a two armed stellar potential. This potential is somewhat more complicated and represents the spiral arms as a superposition of oblate spheroids \citep{1956BAN....13...15S} whose loci are placed along a modified logarithmic spiral function. Each of the spheroids themselves have an linear internal density profile of $\rho_{ss}(a,R)=p_0(a,R)+ap_1(a,R)$ where $a$ is the distance from their centre. The authors suggest that the density parameters $p_0(R)$ and $p_1(R)$ themselves follow either a linear or logarithmic decay with increasing distance, $R$, from the Galactic centre. They find the logarithmic decrease and lower arm mass is most effective at creating secondary arm structures in the gas, so we adopt the same here. All of our arm models are assumed to be logarithmic, with constant pitch angles and are evenly spaced azimuthally.

\vspace{5mm}
We have included in essence two different potential prescriptions for the arms and bar. The WK and CG potentials are sinusoidal perturbations of the axisymmetric disc, whereas the LM and PM arms add an extra mass component to the stellar system.
\subsection{Constructing \lv emission maps}
\label{makinglvmaps}

\subsubsection{Radiative transfer \lv maps}
Rather than simply making kinematic \lv maps of the Milky Way as in previous studies of Galactic structure, we utilise a 3D radiative transfer code to produce synthetic emission maps to compare with observational data. We use a 3D adaptive mesh refinement (AMR) grid radiative transfer code, \torus\, \citep{Harries11072000}. \torus\, is capable of creating synthetic brightness temperature, $T_B$, data cubes of the CO (J=0-1) transition, enabling us to compare our simulations directly with the map of \citet{2001ApJ...547..792D}. \torus\, has been employed in several studies already to create synthetic emission from SPH simulations. Synthetic HI maps of the spiral galaxies of M31 and M33 were created by \citet{2010MNRAS.406.1460A}, finding good agreement with observed emission. \citet{2010MNRAS.407..405D} and \citet{2012MNRAS.422..241A} also used \torus\, to create synthetic emission maps of a subsection of our Galaxy. 

Molecular material traces out the global spiral structure of galaxies (e.g. M51; \citealt{2013ApJ...779...42S}, various external galaxies; \citealt{2003ApJS..145..259H}, and the review of \citealt{1991ARA&A..29..581Y}), and can appear as clear structures in \lv maps such as that in Fig.\,\ref{DameLV}. CO has the added advantage of having a much higher arm-interarm contrast than HI, which is present throughout the Galactic plane \citep{1986ApJ...305..892D,1987ApJ...315..122G}. As such, we choose to primarily concentrate our efforts on reproducing the CO distribution, rather than HI.

The procedure to create $l$-$b$-$v$ data cubes, analogous to those created from observations, is described in detail in \citet{2010MNRAS.406.1460A} and we will give only a brief description here. The SPH data must first be converted to a grid for use by \torus. The grid is filled with SPH particles using an octree method, where the grid is initially a $2\times 2 \times 2 $ cube. The grid is then subdivided according to a mass per unit cell criterion, thereby providing greater refinement in regions of high particle concentration. Our grid is somewhat larger than previous works of \citet{2010MNRAS.407..405D} and \citet{2012MNRAS.422..241A} that focused on the second quadrant alone. As such, to make the grid manageable in terms of memory and map construction time, we use a higher mass per unit cell of $4\times 10^{4}\,\rm M_\odot$ where each particle has a mass of $1.6\times 10^{3}\,\rm M_\odot$, giving approximately 25 particles per cell. We find that lower mass thresholds have very minimal effects on the resulting \lv maps. 
The SPH particle properties including HI and CO fractions, temperature and velocities are mapped on to the grid using a summation of SPH kernels with a Gaussian form. The opacity and emissivity, assuming local thermodynamic equilibrium, are then calculated and stored in the AMR grid for use in the radiative transfer ray-tracing. 

The ray-trace is then performed with input values for the observer coordinates, requiring the distance from the Galactic centre, $R_{\rm obs}$, the azimuthal position in the disc, $\l_{\rm obs}$ and the circular velocity, $V_{\rm obs}$. For a certain velocity channel rays are propagated from the observer throughout the disc in a range of $0^\circ<l<360^\circ$ and $|b|<6^\circ$. While out of plane emission is of minor importance for studying the Galactic disc, we pass rays out of the plane in a high enough latitude so we can produce an integrated emission map of comparable strength to that of \citet{2001ApJ...547..792D}. As a ray enters a cell the intensity of emission is updated from $I_\nu$ to $I'_\nu$ using the opacity, emissivity and optical depth of the current cell at the frequency of interest $\nu$ ($\epsilon_\nu$, $\kappa_\nu$ and $d\tau$ respectively) via;
\begin{equation}
I'_\nu = I_\nu e^{-d\tau}+ \frac{\epsilon_\nu}{\kappa_\nu} \left( 1- e^{-d \tau} \right),
\end{equation}
allowing for the optically thick or thin treatment of the CO (J=0-1) transition, taking full account of optical depth effects.
The intensity is then transformed into brightness temperature by using the Rayleigh-Jeans approximation, $T_B=I_\nu \lambda^2/2 k_B$. This process is then repeated for each velocity channel of interest, resulting in a cube of $T_B$ as a function $l$, $b$ and $v_{los}$. The data cube is then integrated over the latitude dimension ($|b|<2^\circ$) to produce an \lv map analogous to that in Fig. \ref{DameLV}. The number of velocity channels is considerably higher in the central galaxy in order to encompass emission up to a maximum of 280\kms\, seen in the CO observations. To avoid passing rays through empty regions of \lv space we use a number of channels that varies as a function of longitude, tailored to encompass the emission seen in Fig.\,\ref{DameLV}.

\subsubsection{Kinematic \lv maps and the observer coordinates}
\label{KinematicMapping}
When building synthetic \lv maps there is another substantial parameter space that needs to be explored, the coordinates of the observer ($V_{\rm obs}$, $R_{\rm obs}$ and $l_{\rm obs}$). The International Astronomical Union (IAU) recommends values for $V_{\rm obs}$ and $R_{\rm obs}$ of 220\,\kms\,and 8.5\,kpc respectively, but there are a wealth of other values used in the literature (see \citealt{1993ARA&A..31..345R} and \citealt{2008IAUS..248..450M} and references therein). The choice of these parameters has a large effect on the \lv map constructed from simulations. A shift in an observers position of only 0.5\,kpc could make the difference between a spiral arm lying in the inner or outer galaxy, completely altering its position in \lv space.

We fit each simulated CO \lv map  to the observed map to find a best fit $V_{\rm obs}$, $l_{\rm obs}$ and $R_{\rm obs}$. In order to fully explore the observer parameter space we would need to construct numerous \lv maps. If we were to construct full radiative transfer maps for each point in the observer parameter space the computational cost would be extremely high as this would have to be done for each model, at each time step of interest. We instead use approximate \lv plots to fit to the observers position, rather than performing radiative transfer calculations for each observer position. By doing so the computational time is reduced from the order of a day to seconds to build a single CO \lv plot, allowing a fast sweep though observer coordinates. Once the best-fitting observer position is known for a specific galactic simulation we then perform a full radiative transfer calculation with \torus\, to construct a map that is used for comparing different spiral/bar models.

These purely chemo-kinematically derived maps are a simplification compared to those constructed with \torus, but give a good idea of the position of the emission in \lv space, and a rough idea of its intensity.
The maps are constructed as follows. First we choose the observer coordinates from the grid of observer parameter space ($V_{\rm obs}$, $l_{\rm obs}$ and $R_{\rm obs}$). Then we calculate a synthetic CO brightness temperature, $T_{B,\rm synth}$, from each SPH particle in the simulation using the particle's velocity, position and chemical abundance (which is heavily density dependent). To do this, we use a simple radiative scaling law of the form
\begin{equation}
I_{i,\rm synth} \propto \chi_{i,\rm CO}  / d^m_i
\label{synthLV}
\end{equation}
where $\chi_{i,\rm CO}$ is the abundance of CO for the $i^{th}$ SPH particle and $d_i$ is the distance from the observer to the particle. We have tested numerous values for the $m$ parameter and find that $m=2$ gives a synthetic emission map that is very similar to the actual emission map built by the \torus\, radiative transfer code (see Fig.\,\ref{synthEmEG}). This is similar to the approach of \citet{2012MNRAS.421.2940D}, except we do not need to assume the density profile of the ISM gas as it is provided by the SPH particles in the abundance of CO. While the brightness temperature does not technically follow an inverse square law, the column density (and therefore opacity) of material the emission passes through does increase with distance. The $I_{\rm synth}$ factor is then scaled for each particle to match the range of emission in the observed CO map, giving a value of $T_{B,\rm synth}$ for each particle. 

A longitude velocity map is then constructed using the SPH particle coordinates and assuming the observer is on a purely circular orbit. The particles are all first rotated by $l_{\rm obs}$ and then their line-of-sight velocity is calculated as given in \citet{1987gady.book.....B};
\begin{equation}
v_{{\rm los},i} = \sqrt{  v_{x,i}^2 + v_{y,i}^2 } \sin{(l_i -\theta_{v,i})} - V_{\rm obs}\sin(l_i),
\label{Vlos}
\end{equation}
where simple geometry gives the longitude of the particles, $l_i = \arctan(y_i-y_{R_o}/x_i-x_{R_o})$ and the velocity vector is at an angle of $\theta_{v,i}=\arctan(v_{y,i}/v_{x,i})$. An extra $b$ factor for latitude dependence can also be included but it made no difference to the quality of the fit, likely because our simulations vary little in the vertical direction. There is evidence that the sun exhibits peculiar motion relative to the local standard of rest. We investigated adding peculiar motion (up to 20\kms) for a single model and the resulting best-fitting map showed little difference to the case of a circular orbit. For the remainder of the paper, we assume circular orbits to reduce our parameter space.

The emission (in log-brightness temperature) of the particles is then binned into a grid of \lv space of the same resolution as the \citet{2001ApJ...547..792D} CO map (0.125\arcdeg\,by 1\kms). Particles act as a point source, with emission occupying a single pixel of \lv space. To better represent the structure of ISM observations the emission from each particle is broadened and smoothed out into neighbouring \lv bins. The intensity was smoothed using a Gaussian profile with a half width of 1.125\arcdeg\, in longitude and 4\kms\, in velocity. These broadening parameters, as well as the $m$ factor in equation \ref{synthLV} were determined by fitting to an \lv map built by \torus. The 4\kms\,velocity smoothing matches the turbulent velocity width we add to the \torus\,maps (discussed in section \ref{Results}). No additional smoothing in longitude is added to the \torus\, maps as the grid-cell structure of the code ensures emission comes from sources of finite size.

These simple maps enable us to find a best-fitting map for each individual Galactic simulation, removing the uncertainty in placing the observer at some arbitrary position. The range of observer coordinates investigated in this fit are given in Table \ref{POTparams}. Once a best fit is known, \torus\, is then used to build a full map using the best-fitting observer coordinates, which can then be used to compare the different galactic potentials.

To quantify the goodness of fit for each model we use a simple fit statistic. We calculate a mean absolute error (MAE) in log-$T_B$ between the synthetic map and the CO map of \citet{2001ApJ...547..792D}, shown in Fig.\,\ref{DameLV}. This is then normalised by the number of pixels with non-negligible emission in the observed \lv map, $n_{pixels}$, to obtain a fit statistic close to unity. The form of our fit statistic is thus,
\begin{equation}
{\rm Fit} = \frac{\sum_{pixels} |T_{B,synth}d b - T_{B,{\rm Dame}}d b|}{n_{pixels}} {\rm 1\,K^{-1}\,arcdeg^{-1}},
\end{equation}
where $T_B  db$ is the brightness temperature integrated over latitude, and ${\rm 1\,K^{-1}\,arcdeg^{-1}}$ ensures a dimensionless statistic.
Our choice of MAE over RMS is to ensure that single pixels far from the observed value do not cause severe deterioration in the fit statistic, as we are interested in a global match, rather than whether a individual features can be exactly reproduced. Because our simple approximate \lv maps and those made using radiative transfer are calculated using two very different methods the fit statistic should not be quantitatively compared between these two different types of map. However, the relative strength of emission features, and the general morphology, can be.

\section{results}

\subsection{General features of radiative transfer maps}
\label{Results}

\begin{figure}
 \includegraphics[trim = 10mm 0mm 0mm 0mm,width=84mm]{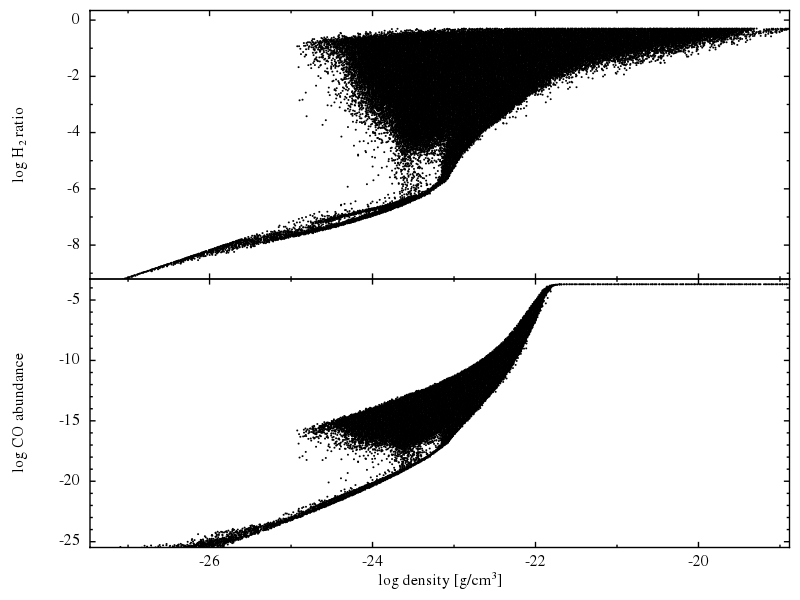}
  \caption{The \H2 ratio (top) and CO abundance (bottom) verses individual SPH particle density for a simple 4-armed disc galaxy simulation of 1\,million particles after 200\,Myr of evolution.}
  \label{Chem}
\end{figure}

The abundance of \H2 and CO as a function of particle density is shown in Fig. \ref{Chem}, from a simple 4-armed disc galaxy simulation of 1 million particles after 200\,Myr of evolution. The particles perform loops in abundance-density space as they pass into and out of high density regions (for an in-depth discussion see \citealt{2008MNRAS.389.1097D}). The CO plateaus at the abundance of C used in the cooling routines and CO chemistry ($2\times 10^{-4}$), which is not evolved in our simulations.
The CO and density evolution reaches a steady state after approximately 300\,Myr. We choose to run our simulations until at least 354\,Myr and until a maximum of 472\,Myr\footnote{The evolution times frequently used in this paper of 236, 354 and 472 Myr correspond to 1/2, 3/4 and 1 times 10 code units, determined from the astronomical constants: $G$, $M_\odot$ and kpc.}. Arm and bar structures are also well developed by these times, but will continue to slowly evolve on the order of Gyr.

\begin{figure}
 \includegraphics[trim = 25mm 0mm 0mm 0mm,width=92mm]{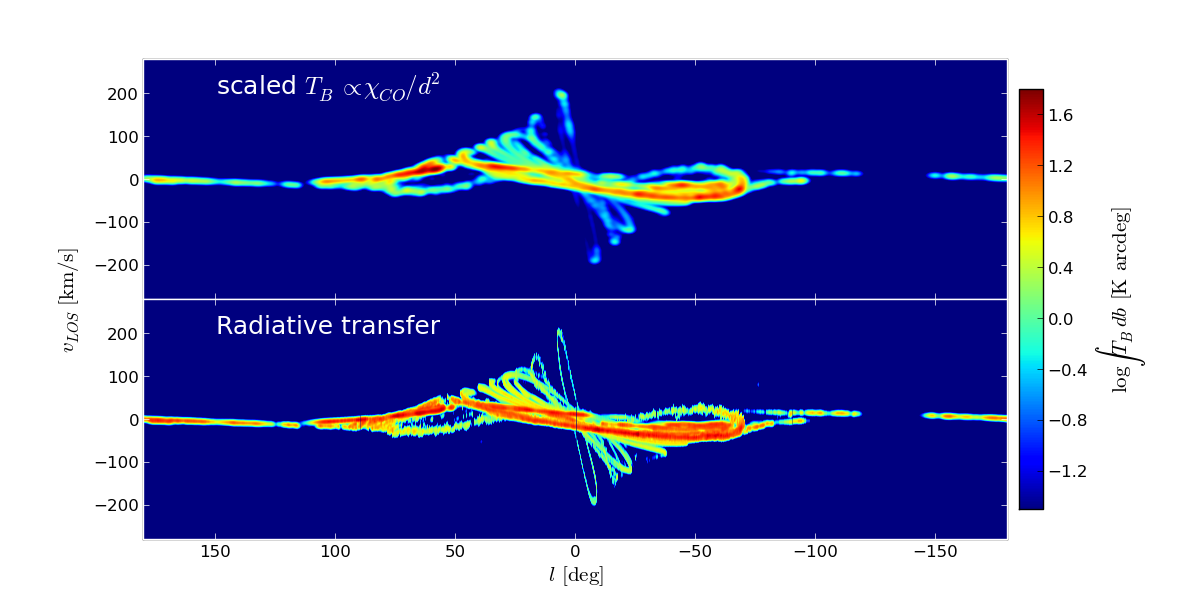}
  \caption{Two emission maps of a barred Milky Way simulation, of WK type. Top: synthetic CO \lv map constructed using equation \ref{synthLV}. Bottom: \lv map created at the same observer coordinates but with the radiative transfer code \torus, with the same values of $R_{\rm obs}$, $V_{\rm obs}$ and $l_{\rm obs}$.}
  \label{synthEmEG}
\end{figure}

Figure \ref{synthEmEG} shows \lv maps constructed within a barred galaxy simulation. In the upper panel we show a map made using the method described in Section \ref{KinematicMapping}. The lower panel shows a map made using \torus. Both are constructed using the same values for \Vobs, \Robs\, and \lobs. Both maps trace the same regions of \lv space, with roughly the same intensities. The simple map underestimates the emission in some regions, and overestimates in others. This is expected; if the simple map reproduced the \torus\, map there would be no need to perform the radiative transfer calculation.

The intensity out of the plane is integrated through $\pm 2^\circ$ to match the map produced in \citet{2001ApJ...547..792D}. The integration usually does not introduce any new features in \lv space as our simulations are effectively confined to the Galactic plane. The contrast between emission in our \torus \, maps is comparable to that of the observations in the inner Galaxy (Fig.\,\ref{DameLV}). The distribution of emission in general is smoother than that seen in observations. This is a result of the continuous nature of the potentials, which are idealised compared to the arm structures in observed spiral galaxies.

Early tests using \torus\, for CO \lv maps showed that the features created were far too narrow in velocity width compared to observations. To resolve this, we added a turbulent velocity to the width of the CO line emission profile of 4\kms, a value high enough to smear out the fine emission features but not so strong as to blend features in \lv space (see Fig.\,\ref{ResTest} in Appendix \ref{ResStudy} for an example without a turbulent velocity). This is at the lower end of ranges suggested by CO observations of the outer regions of disc galaxies (see \citealt{2006ApJ...638..797D} and references therein). The turbulent velocity could be scaled as a function of some cloud size determined by the clumpiness of SPH particles \citep{1981MNRAS.194..809L,1986ApJ...305..892D,2004RvMP...76..125M}. However, we choose a constant factor to avoid introducing additional variables. Quantitative tests of the chemistry and radiative transfer in CO, including detailed comparisons to observations, will be the subject of a future study (Duarte-Cabral et al., in preparation).

The strength of the CO emission in our \torus\, maps is somewhat higher than that observed, peaking at approximately 40K compared to 20K seen in observations. The contrast between the synthetic and observed \lv maps is however similar when a turbulent velocity of 4\kms\, is included. There are several possible reasons for this difference. The first is that the strength of the CO emission is very sensitive to the surface density of the ISM disc. We performed initial simulations using half the mass of gas used here ($4\times 10^{9}\rm M_\odot$) and emission from arm/bar features was very weak in \lv space. The disc mass found through integration of the disc surface density profile resulted in visible emission from the arm features, and so was used for the simulations presented here. Another consideration is that the production of CO has no limit other than the maximum amount of C allowed to be present in the ISM. All SPH particles tend to increase their molecular abundance (and density) up to this limit, as there is no process to break up and heat the gas. Additional heating mechanisms such as stellar feedback or magnetic fields would be required to break up the dense clumps of ISM gas and remove some of the excess CO build up. The addition of stellar feedback would also cause material to be more dispersed vertically compared to the no feedback case \citep{2006ApJ...641..878T,2011MNRAS.417.1318D,2012MNRAS.422..241A}. The gas in the simulations shown here is very confined to the \xy plane, as there is no mechanism to drive the gas off-plane and counteract the disc potential. This causes all the molecular material to be within a single latitude channel in the construction of the emission data cubes, increasing the strength of emission seen in \lv space. Conversely, there is a considerable amount of off-plane emission seen in observations \citep{1987ApJ...315..122G,1990A&A...233..437B,2001ApJ...547..792D}.

We begin by first modelling bar and arm potentials separately (Sections \ref{baronly} and \ref{ArmSec}). Using theses results we then combine our best-fitting arm and bar potentials in Section \ref{MixModels}.

\subsection{Bar only simulations}
\label{baronly}

\subsubsection{Simulation \xy maps}
\begin{figure*}
\includegraphics[trim = 15mm 0mm 0mm 0mm,width=185mm]{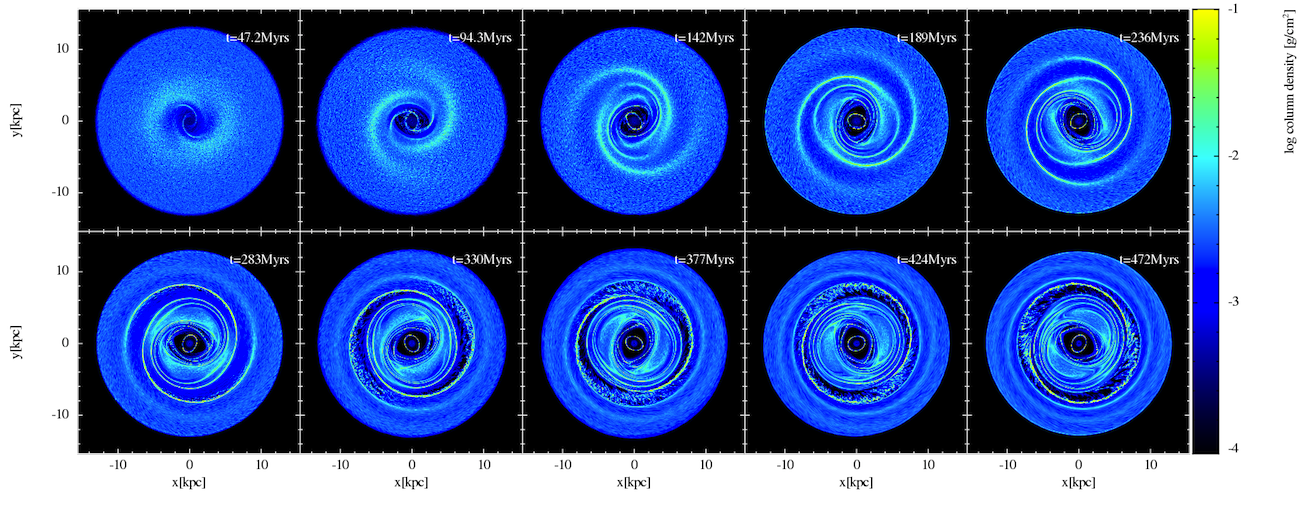}
\caption{The evolution of the bar model of \citet{2001PASJ...53.1163W} with a core radius of 2\,kpc rotating at 50\,\ps. Both the bar and the gas are rotating clockwise. Note that the morphology is effectively the same from 350-470\,Myr and the arms will eventually wind up to form a ring-like structures with elliptical orbits parallel and perpendicular to the bar major-axis. The orientation of the features in this and other top-down figures is determined by the initial alignment of the non-axismmetric potential with the x-axis at $t=0$Myr, orientation does not correspond with any of the \lv maps shown in other figures.}
\label{BarredEG}
\end{figure*}

An example of the evolution of a barred galaxy simulation is shown in Fig.\,\ref{BarredEG}, using the bar model of \citet{2001PASJ...53.1163W} with a core radius of 2\,kpc rotating at 50\,\ps. When using different bar potentials the overall evolution is similar. The bar potential is active throughout the entire simulation, and gas within the bar establishes elliptical orbits along the major axis of the bar from 100\,Myr onwards. After 150\,Myr the gas in the outer disc displays a two armed spiral structure inside the OLR, the strength of which is related to the core radius and strength of the potential. These arms are not in a steady state, and their pitch angle is decreasing over time. After about 4 rotations of the bar (the last panel in Fig.\,\ref{BarredEG}) the arms are wound up enough that they begin to join to create elliptical/ring-like structure at the OLR, with the orbits set as being either perpendicular to the bar inside the OLR or parallel to outside the OLR \citep{1995gaco.book.....C,1996FCPh...17...95B}. Any arm potential we combine with these bars would be substantially subdued in this region, which is near to the solar radius.

In test calculations where we use an isothermal equation of state to model the ISM the arms driven by the bar are maintained when the temperature is high (10000K). However, in low temperature isothermal cases and adiabatic+cooling cases the arms enclose on the aforementioned set of orbits around the OLR. There also exists a set of orbits perpendicular to the bar in the inner galaxy. These orbits (commonly referred to as $x_2$ orbits) only exist when there is a region between two separate ILR's \citep{1980A&A....92...33C}. In calculations where we used a more simplified axisymmetric potential (a bulge-less flat logarithmic potential) there were no such inner orbits as there was only a single ILR. However, the rotation curve we use here has an inner bulge (see Fig.\,\ref{MWLR}), providing a second ILR and so setting up a family of inner perpendicular orbits. These orbits are seen in other works using analytic barred potentials (e.g. \citealt{2009AstL...35..609M}).

\begin{figure*}
\includegraphics[trim = 15mm 100mm 0mm 0mm,width=185mm]{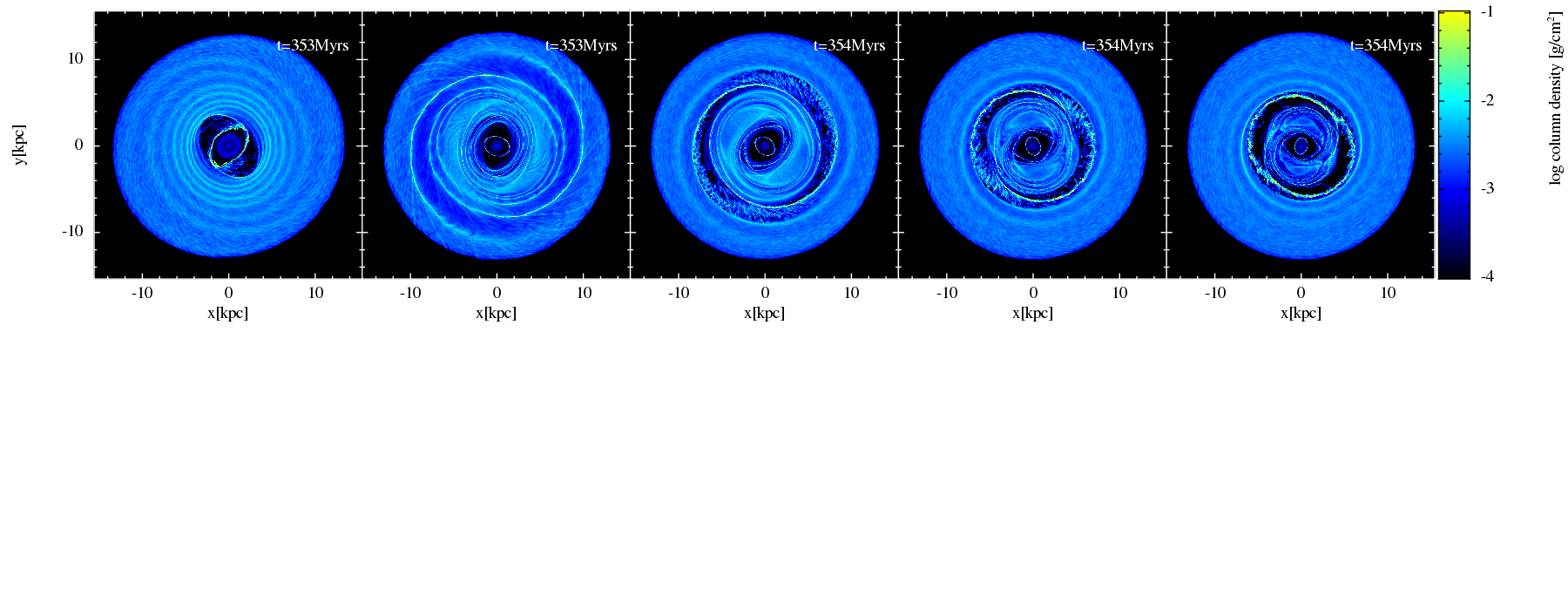}
\caption{The bar model of \citet{2001PASJ...53.1163W} with a core radius of 2\,kpc rotating at pattern speeds of 20, 40, 50, 60 and 70\,\ps\,, increasing from left to right, at a time of 354\,Myr. The gas and potentials are rotating clockwise viewed from above. These top-down maps correspond to the central row of Fig.\,\ref{WDBarPatternLV}. The contraction of the outer Lindblad resonance is clearly as $\Omega_b=50\rightarrow 70$\ps.}
\label{WDBarPSxy}
\end{figure*}

The pattern speed of the bar is key in determining the structures that develop in the inner galaxy. Plots of  the WK bar model are shown at various pattern speeds in Fig.\,\ref{WDBarPSxy}. All the bar potentials used in this study display similar behaviour as a function of pattern speed. As the pattern speed increases, the ILR and OLR contract, reducing the radial extent of features driven by the bar. There is also an inability of the slower bars to drive any strong arm-like features compared to the faster pattern speeds, owing to the fact that the OLR is beyond the edge of the Galactic disc. The slower bars also have a greater impact on the dispersion in the rotation curve compared to the faster bars. The 20\,\ps\,bar in Fig.\,\ref{WDBarPSxy} has a dispersion of around $\pm 50$\kms\,at $R=2$kpc. Conversely, the faster bars have a greater variation in the rotation curve in the outer regions of the disc corresponding to the location of the driven arms, but of a much smaller scale than that of the inner region of the slow bar.

\begin{figure}
\includegraphics[trim = -30mm 0mm 60mm 0mm,width=70mm]{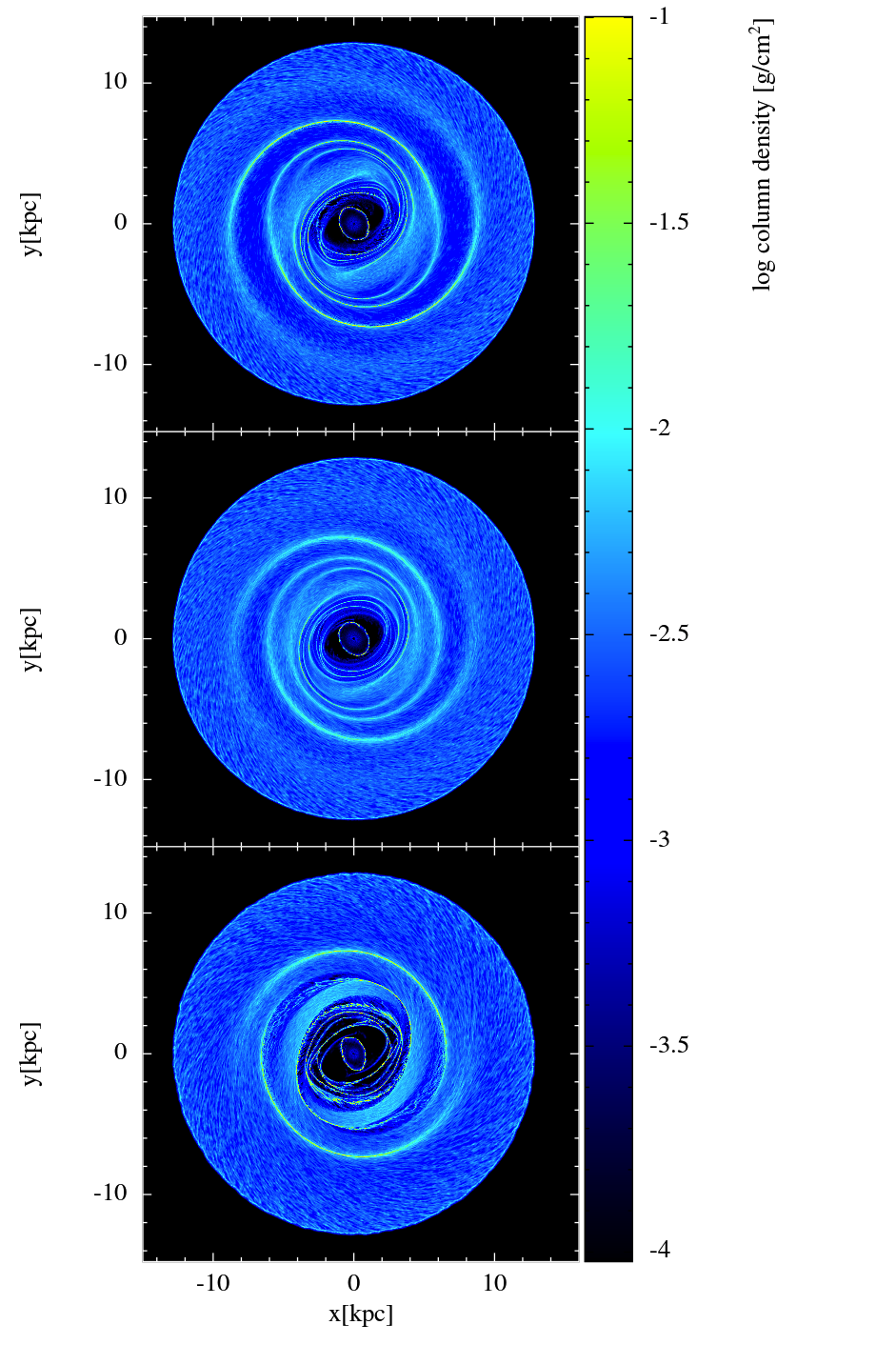}
\caption{Different bar models angled at 45\arcdeg to the Sun-Galactic centre line with a pattern speed of 50\ps after 236\,Myr of evolution. The models (top to bottom); WK, WKr2 and LM are described in the main text.}
\label{BarStrength}
\end{figure}

Figure \ref{BarStrength} shows a comparison between our three different bar models. All have a pattern speed of 50\ps\, and are shown after 236\,Myr of evolution angled at 45\arcdeg with respect to the Sun-Galactic centre line. The inner structure is similar for all models. Immediately outside this there are other thin orbital structures, more so in the case of the LM bar. The arm structures generated in the outer disc are different in each model. The LM bar has formed very tightly wound arms compared to the others, a result of the different radial drop-off compared to the other models. The LM bar potential is thinner along the semiminor axis than the others, which could also contribute to the tighter arm structures. The WK and WKr2 bars differ in the extent of their central core radius, the effect of which can be seen in Fig.\,\ref{BarStrength}. The bar with the smaller core radius has weaker arms compared to the bar with a larger core.

\subsubsection{Kinematic and radiative transfer \lv maps}
\begin{figure}
 \includegraphics[trim = 25mm 0mm 0mm 0mm,width=88mm]{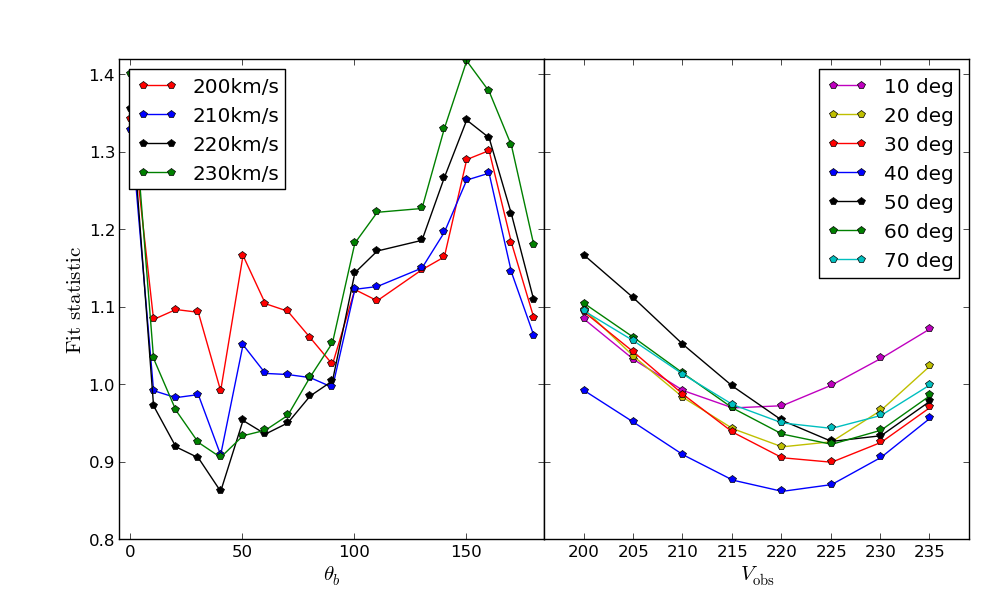}
  \caption{An example of fitting to the observer's coordinates using simplified \lv maps as described in Section \ref{makinglvmaps}. Here we show the fit statistic for a barred Milky Way after 470\,Myr of evolution. The fits to the observers azimuthal position and circular velocity are shown in the left and right panels respectively at $R_{\rm obs}=8.5$\,kpc (the fit as a function of $R_{\rm obs}$ is not shown for clarity). The different coloured lines show the fit for a certain value of $V_{\rm obs}$ (left) or $\theta_b$ (right).}
  \label{PrefitEG}
\end{figure}

An example of the results of fitting to the observer's coordinates is shown in Figure \ref{PrefitEG}. The galaxy model used in this example is a WK barred galaxy with a bar pattern speed of 50\,\ps. The parameter sweep is performed at a timestamp of 470\,Myr and the bar major-axis lies along the y-axis by default. The left-hand panel of Fig.\,\ref{PrefitEG} shows that a best fit orientation of $\theta_b=40^\circ$ is preferred, broadly in keeping within the accepted range. The fit as a function of velocity gives the IAU standard value of 220\kms, but it is clear the velocity fit is not as well constrained as the bar orientation. While not shown here, $R_{\rm obs}$ is similar to $V_{\rm obs}$ in that it shows a shallow global minimum. This is the case with most potential models, with the $l_{\rm obs}$ parameter showing the clearest troughs/peaks of the fit statistic. The $l_{obs}$ parameter is only shown between 0-180\arcdeg, as the potentials, and fit statistic, are symmetric.

\begin{figure*}
\includegraphics[trim =30mm 10mm 0mm 0mm,width=180mm]{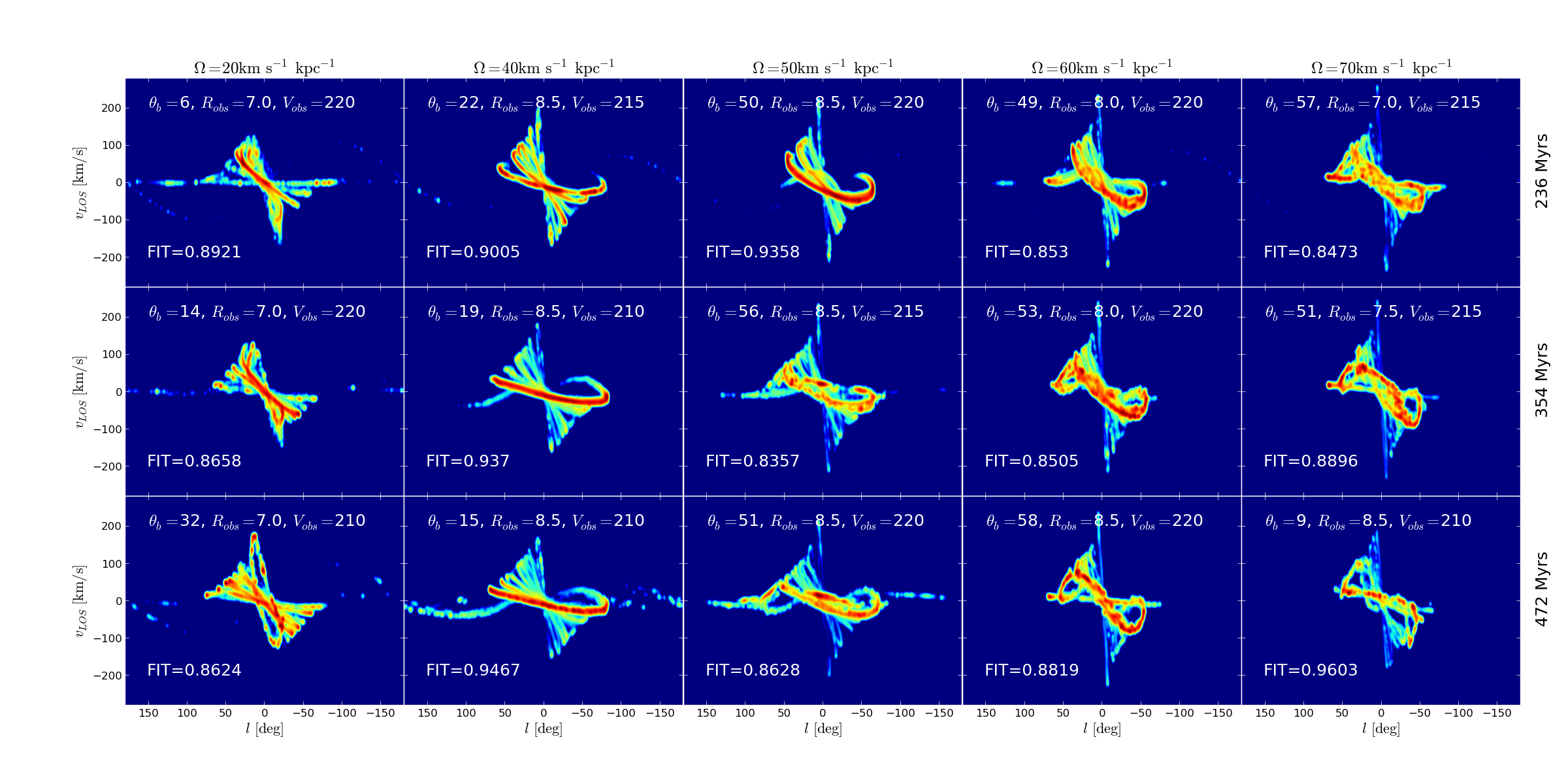}
\caption{The bar model of \citet{2001PASJ...53.1163W} with a core radius of 2\,kpc rotating at pattern speeds of 20, 40, 50, 60 and 70\,\ps\, increasing from left to right with time increasing from top to bottom (236, 354, 472\,Myr). The values for the bar orientation, observer distance and circular velocity, and fit statistic are overplotted on each $\Omega_b$-$t$ pair (in degrees, kpc and \kms\,respectively).}
\label{WDBarPatternLV}
\end{figure*}

\begin{figure}
\includegraphics[trim = 0mm 0mm 0mm 0mm,width=84mm]{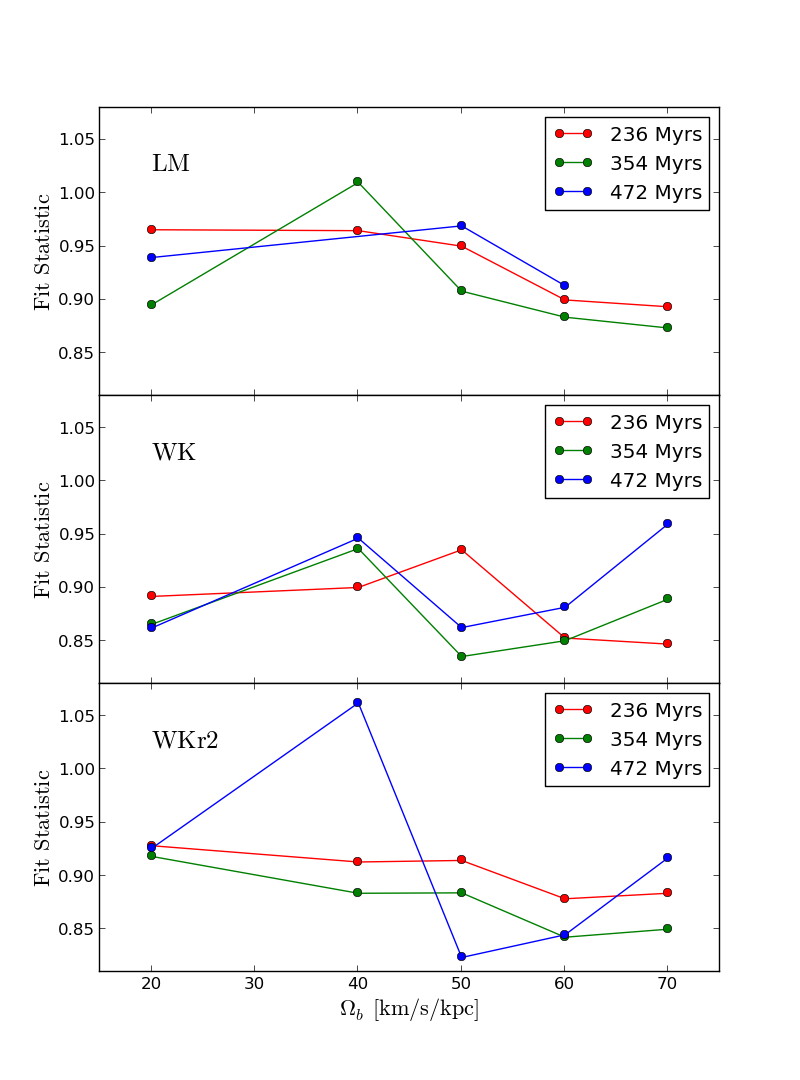}
\caption{The fit to pattern speed across all bar models. There is a slight preference towards 50-60\,\ps. Note that the LM bar has a poorer fit statistic overall, and that the simulations of this bar were halted before it reached the final timestamp for $\Omega_b=40$ and 70\,\ps.}
\label{BarPattenFit}
\end{figure}

The results from the fit to all bar parameters ($\Omega_b$, $\theta_b$, \Robs\, and \Vobs) are shown in Fig.\,\ref{WDBarPatternLV}. These \lv maps are from the simulations shown in Fig.\,\ref{WDBarPSxy}, and are constructed using the method outlined in Section \ref{KinematicMapping}. We do not show the maps of the WKr2 and LM bars but include their best fit results in Fig.\,\ref{BarPattenFit} and Table \ref{BarResults}.  Fig.\,\ref{WDBarPatternLV} shows the best-fitting \lv plots for pattern speeds of 20, 40, 50, 60 and 70\ps\,at 236, 354 and 472\,Myr of evolution. The parameters for each of the best-fitting maps ($\theta_b$, $R_{\rm obs}$ and $V_{\rm obs}$) are overplotted on to each individual map, along with the corresponding fit statistic. The orientation of the bar with respect to the Sun-Galactic centre line is effectively a free parameter in our fitting to the observer coordinates. 

Inspection of Figure \ref{WDBarPatternLV} shows that bars moving at 50-70\ps\,tend to favour an orientation of around 50\arcdeg, while the lower pattern speeds favour lower values. This is a result of the shift in the OLR from the external Galaxy to the internal Galaxy as we increase pattern speeds, and the resulting location of the arms driven by the bar. For lower pattern speeds the arms extend to outside the solar radius, up to the OLR. This means these arms fit the outer quadrants, while the central bar structure fits the inner quadrants. For the higher pattern speeds the driven arm structures lie inside the solar radius, and so the bar and arm structure is contained within the inner Galactic quadrants alone, leaving the outer quadrants empty. The resulting two different bar pattern speed domains cause the different bar orientation ranges. Our grid of values for the $\theta_b$ parameter is fairly coarse, incrementing in steps of 10\arcdeg\, from the bar's position at times of 236, 354 and 472\,Myr after being initially aligned with the x-axis at $t=0$Myr. As such there is an uncertainty up to 10\arcdeg\,in the values given here. This means that the frequently used value suggested for the ``Long bar" by \citealt{2009PASP..121..213C} of $\theta_b=45^\circ$, is within the bounds of the values found here by our best-fitting bars with $\Omega_b=$50-60\ps.

The \lv maps shown in Figure \ref{WDBarPatternLV} rarely generate considerable structure in the outer quadrants. The exception is the 40\ps\,model at later times, where the arm structures driven by the bar persist into the outer disc due to the OLR's position beyond the solar radius. At later times the arm structures driven by the bar join to create closed orbits, that are clearly visible in the \lv diagram (especially for the $\Omega_b=70$\ps\,cases). While not shown here, the \lv maps of the WKr2 bar are very similar morphologically.

The best-fitting structures fit one of two regions well. The first category of good fitting maps are those that simply fill out more structure in \lv space, such as the 50\ps\,WK bar at 472\,Myr (bottom central panel of Fig.\,\ref{WDBarPatternLV}). In these cases the arms driven by the bar extend to relatively large radii, spreading the emission into a larger range of $|l|$. The other category of good fits are those where the strength of the emission in the inner Galaxy follows a pattern similar to the observed CO map. This ridge of CO emission not present in HI is often attributed to a molecular ring-like structure, but could also be explained by arm or bar features of the correct geometry (\citealt{2001ApJ...547..792D}, \citealt{2012MNRAS.421.2940D}). In Fig.\,\ref{WDBarPatternLV} at early times, the 60\ps\,bar is a good fit for central emission due to arm-like structures extending to a radius of about 5\,kpc, with a fairly wide pitch angle. By 472\,Myr the arms have closed upon each other, creating an elliptical structure where the arms once were. Both early and late times fill out the same area of \lv space, but the advantages of an arm structure over that from a ring is that it can curve in the correct direction in \lv space. An elliptical or ring like structure would show twofold rationally symmetry about 0\arcdeg- 0\kms, not seen in the CO data in Fig.\,\ref{DameLV}. The strong central ridge in seen in the 20\ps\, \lv maps in Fig.\,\ref{WDBarPatternLV} seems to provide a reasonable match for the central ridge in the CO data. This structure actually results from the concentric rings surrounding the bar, as seen in Fig.\,\ref{WDBarPSxy}. The addition of an arm potential disrupts these relatively weak structures easily, and are needed to drive outer arm features absent in the 20\ps\,bar. The emission for this bar is also relatively confined to this ridge, in comparison to the early time 60 or 70\ps\, maps.

The \lv maps in Fig.\,\ref{WDBarPatternLV} seem to be heavily time-dependent. Over a 200\,Myr time frame the emission structures can change considerably. The 60\ps\, model in particular changes from having an emission ridge comparable to observations to a looped structure that is a poor by-eye match to the CO data. Maps of the WKr2 bar (a smaller core radius) evolve slower than the WK bar, maintaining their features due to the relatively weaker potential. For example the 60\ps\, map at 472\,Myr does not display the strong figure-of-eight like structure seen in the equivalent map of the WK bar (Fig.\,\ref{WDBarPatternLV}).

A comparison of the fit statistic as a function of $\Omega_b$ for all our bar models at the three different timestamps is shown in Fig.\,\ref{BarPattenFit}, and the best-fitting values are explicitly shown in Table \ref{BarResults}. At first glance there seems to be no strong relation between the goodness of fit and $\Omega_b$. There are however some common features between the different models. The 40\ps\,models tend to have some of the worse fits, for reasons discussed above relating to the position of arms in the outer Galaxy. The best-fitting speeds tend to be in the $\Omega_b>40$\ps\, range. The best-fitting pattern speed for the WK and WKr2 bars is 50\ps, though the 60\ps\,is also a comparably good fit. While the 70 and 20\ps\,pattern speeds are numerically a good fit in some instances, we choose to not include these in our models with arm and bar potentials. This is because of the relatively short time-scale on which the \lv emission structure appears a good match to the CO data compared to the 50 and 60\ps\, models. Figure \ref{BarPattenFit} also indicates that overall the LM bar is a poorer fit than the model of \citet{2001PASJ...53.1163W}, so we choose not to follow these up for further analysis in combination with arm potentials. This bar is somewhat thinner than the bar of \citet{2001PASJ...53.1163W} due to our choice of axis ratios. The quality of the fit could be a result of the chosen axis ratios but we do not consider this further.

\begin{table}
 \caption{Best-fitting values for the bar only, arm only, and arm+bar simulations. A systematic uncertainty for each value is present due to the coarseness of the parameter space; $\Delta \Omega_{sp}=\Delta \Omega_b=10$\ps, $\Delta V_{\rm obs}=5$\kms, $\Delta R_{\rm obs}=0.5$kpc and $\Delta \theta_b=10^\circ$. The parameter space for the mix models is smaller than the isolated cases and is discussed in Section \ref{MixModels}.}
 \begin{tabular}{@{}lccc}
  \hline
   & & Bar model & \\
  \hline
  Best-fitting paramater & WK & WKr2 & LM \\
  \hline
  \hline
  $\Omega_b$ [\ps]& 50 & 60 & 70 \\
  $V_{\rm obs}$ [\kms] & 215 & 220 & 235 \\
  $R_{\rm obs}$ [kpc] & 8.5 & 8.5 & 7.0 \\
  $\theta_b$ [\arcdeg] & 56 & 51 & 41 \\
   \hline
   & & Arm model & \\
  \hline
  Best-fitting parameter & & CGN2 & CGN4 \\
  \hline
  \hline
  $\Omega_{sp}$ [\ps]&  & 20 & 20 \\
  $V_{\rm obs}$ [\kms] &  & 210 & 205 \\
  $R_{\rm obs}$ [kpc] &  & 8.0 & 8.5 \\
  $\alpha$ [\arcdeg] &  & 12.5 & 10.0 \\
   \hline
   & & Mix model & \\
  \hline
  Best-fitting parameter & & CGN2+WK& CGN4+WK\\
  \hline
  \hline
  $\Omega_{b}$ [\ps]&  & 50 & 60 \\
  $V_{\rm obs}$ [\kms] &  & 220 & 215 \\
  $R_{\rm obs}$ [kpc] &  & 8.5 & 8.5 \\
  $\alpha$ [\arcdeg] &  & 15 & 10 \\
   \hline
 \end{tabular}
 \label{BarResults}
\end{table}

\begin{figure}
\includegraphics[trim = 0mm 0mm 0mm 0mm,width=84mm]{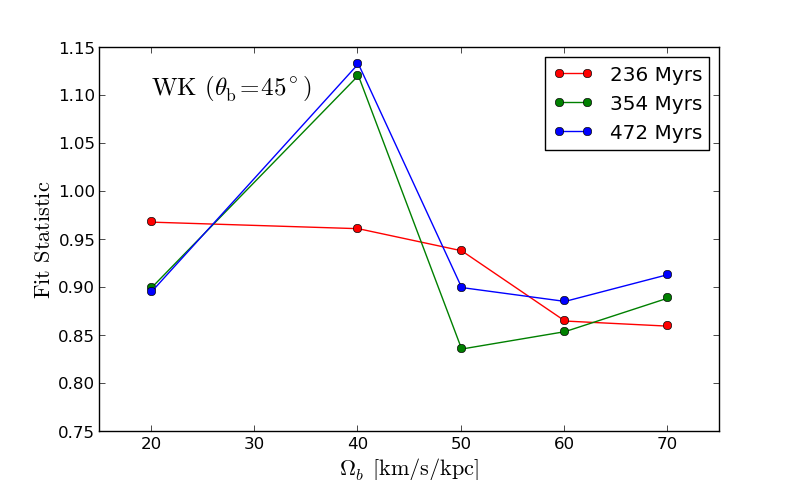}
\caption{The fit statistic for the \citet{2001PASJ...53.1163W} bar when fixed at $\theta_b=45^\circ$ with $V_{\rm obs}$ and $R_{\rm obs}$ left free. The simulations data is identical to that used in Fig.\,\ref{BarPattenFit}.}
\label{Bar45PattenFit}
\end{figure}

Our best-fitting bar models suggest a bar orientation of $\approx45^\circ$, in accordance with observations of the ``Long bar". In Fig.\,\ref{Bar45PattenFit} we show the fit statistic as a function of $\Omega_b$ for the WK bar with $\theta_b$ fixed at 45\arcdeg\,while keeping $V_{\rm obs}$ and $R_{\rm obs}$ free. The lowest fit statistics over all times considered are for the 50 and 60\ps\,models, which is consistent with the fits where $\theta_b$ is left free, and the general trend with $\Omega_b$ is similar to the WK and WKr2 bars in Fig.\,\ref{BarPattenFit}.

\begin{figure}
\includegraphics[trim = 10mm 0mm 0mm 10mm,width=92mm]{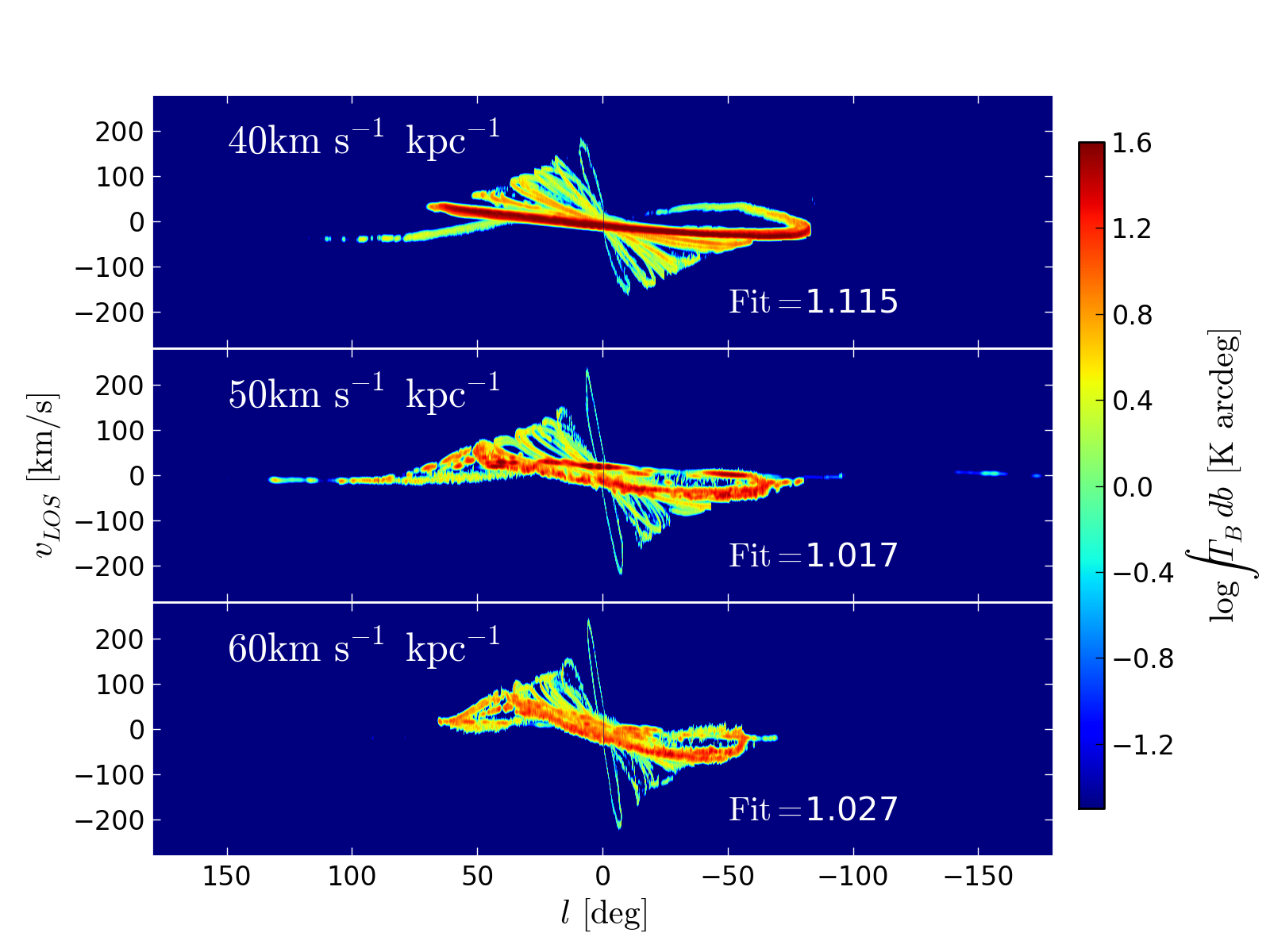}
\caption{Radiative transfer \lv maps constructed using \torus, rather than the simple chemo-kinematic re-mapping method used to create the maps in Fig.\,\ref{WDBarPatternLV}. The bar is that of \citet{2001PASJ...53.1163W} after 354\,Myr of evolution and pattern speeds of 40, 50, 60\ps (increasing from top to bottom). The brightness temperature scale is calculated exactly so the fit statistic is on a different scale to that for the previous maps.}
\label{TorusBarPS}
\end{figure}

We show full radiative transfer maps for only a handful of those models already discussed and we choose to use the radiative transfer to primarily differentiate between full models including bar and arm potentials. The simple $T_B\propto \chi_{\rm CO}/d^2$ maps suffice for narrowing down the wider parameter space. In Figure \ref{TorusBarPS} we show \torus\, maps of the WK bar at pattern speeds of 40, 50 and 60\ps\,after 354\,Myr of evolution. These correspond to the simple maps shown in the centre of Fig.\,\ref{WDBarPatternLV}. The arm feature near the Solar position in the 40\ps\, model is visible as extremely bright emission in the top panel of  Fig.\,\ref{TorusBarPS}. This pattern speed does however provide a better match for the Carina arm feature compared to the faster pattern speeds. As the pattern speed increases, the emission covers a narrower range of longitudes, and increases the line-of-sight velocity of the central emission ridge.
The emission towards the Galactic centre ($|l|<5$) with the greatest $|v_{los}|$ is a very clear feature in the observed CO emission; the CMZ. We find no such strong emission in our maps in Fig.\,\ref{TorusBarPS}. We do see some similar features to the peak velocity structures seen in observations in some of our maps in Fig.\,\ref{WDBarPatternLV}, but there is not enough CO produced to be seen in our \torus\,maps. We discuss this further in section \ref{Discussion}.

We adopt bar pattern speeds of 50 and 60\ps\, to use in our arm-bar mixture models. We chose to run WK bars (which appear stronger in the outer disc) at 50\ps\,and WKr2 bars at 60\ps. This choice is also supported by the fit statistic shown in Fig.\,\ref{BarPattenFit}, which shows that 2/3 of the timestamps investigated have their minima at 50\ps\,for the WK bar and 60\ps\,for the WKr2 bar. We do not follow up the 70\ps\, models because they lose their arm structure relatively fast compared to other models, resulting in ellipses in \lv space. Their speed is also fast enough to sweep up a large quantity of gas inside of 4\,kpc. This would make it impossible for arm structures to exist in the inner Galaxy, making it difficult to see emission not associated with the elliptical bar orbits within $|l|<45^\circ$. We exclude 20\ps\, due to their lack of any arm feature and strong inner resonance features that fail to match the morphology of the inner \lv structure seen in the data. They also lack any inner features that can match the peak velocities seen in the observed CO data. The 40\ps\, models are excluded due to their poorer fit statistics in the case of each model (see Fig.\,\ref{BarPattenFit}).

\subsection{Arm only simulations}
\label{ArmSec}

\subsubsection{Simulation \xy maps}
\begin{figure*}
\includegraphics[trim = 15mm 0mm 0mm 0mm,width=185mm]{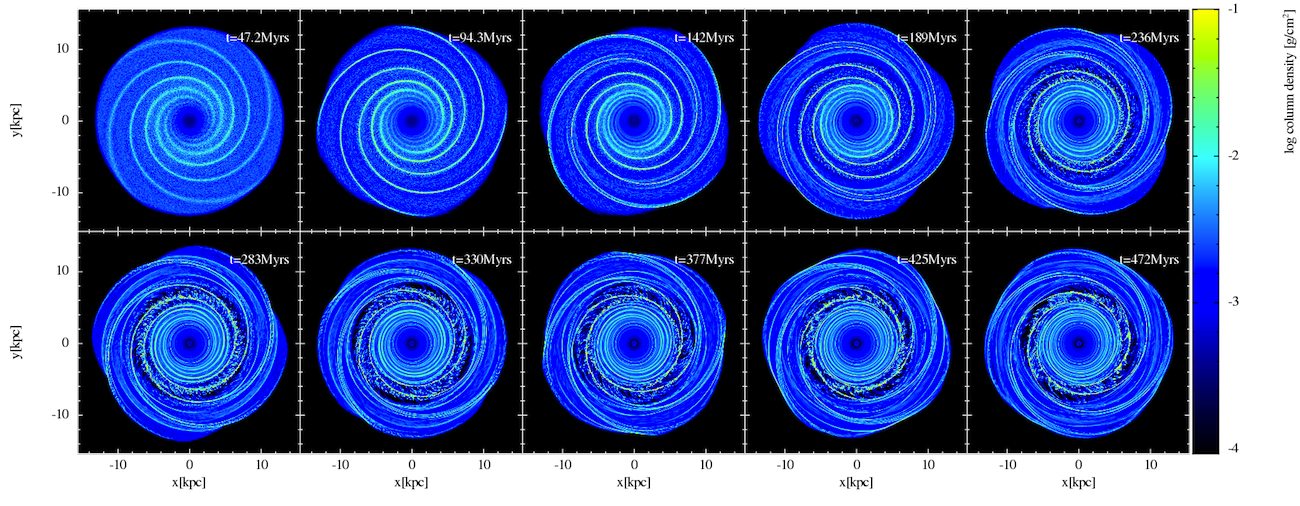}
\caption{The evolution of the 4-armed model of \citet{2002ApJS..142..261C} moving at a pattern speed of 20\ps\,with a pitch angle of 15\arcdeg. Arm spurs are clearly seen near the ILR ($R\approx7$kpc) after 200\,Myr. The outer Lindblad resonance is beyond the simulation radius.}
\label{SpiralEG}
\end{figure*}

An example of the evolution of an isolated CG-type arm model is shown in Fig.\,\ref{SpiralEG}, with the parameters; $N=4$, $\alpha=15^\circ$, $\Omega_{sp}=20$\ps. The spiral structure in the gas tends to survive only between the ILR and OLR region (see Fig.\,\ref{MWLR}), even though the potential is present throughout the disc. For the 20\ps\, case shown in Fig.\,\ref{SpiralEG} the OLR is beyond our simulation radius, but the ILR is clearly seen at later times at $R=7$\,kpc. Around this radius there exists strong spur features as seen in \citet{2006MNRAS.367..873D}. After approximately a Gyr of evolution the gas becomes aligned on 4:1 orbits at the OLR and ILR with spiral arms persisting in between. 

\begin{figure*}
\includegraphics[trim = 15mm 0mm 0mm 0mm,width=185mm]{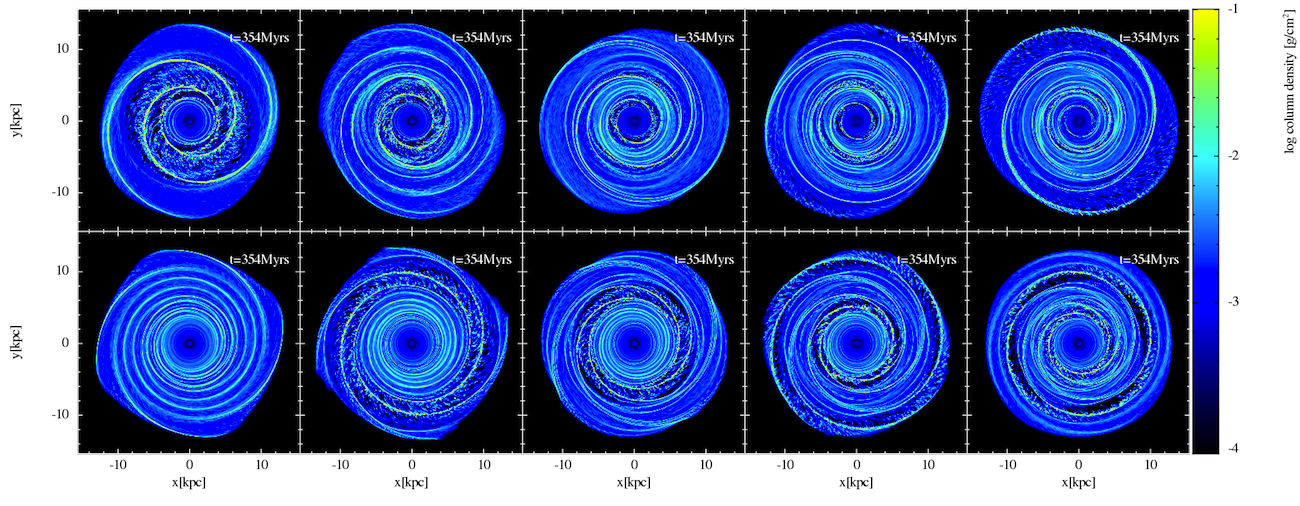}
\caption{Response of the gaseous disc to arm potentials of different pattern speeds. 2-armed and 4-armed models are on the top and bottom rows respectively with increasing pattern speed along the x-axis (10, 15, 20, 25, 30\ps). All models are of that of \citet{2002ApJS..142..261C} after 354\,Myr of evolution with a pitch angle of 12.5\arcdeg.}
\label{CGArmPSxy}
\end{figure*}

A comparison of the ISM gas response to different arm pattern speeds is shown in Fig.\,\ref{CGArmPSxy} for our CG 2 and 4-armed models after 354\,Myr of evolution with a pitch angle of $\alpha=12.5^\circ$. The variation with $\Omega_{sp}$ behaves in a similar fashion for different values of $\alpha$. Each model has a region where spurs exist, the radial position of this decreases with increasing pattern speed and roughly corresponds with the location  of the ILR. Even by-eye it is clear that some of the models in Fig.\,\ref{CGArmPSxy} do not display the desired morphological features. The 10\ps\, $N=4$ models all lacked spiral features that represented the underlying potential. While these models do show spiral structure, the gas is rotating too fast with respect to the potential inside the ILR, resulting in a winding up of spiral features. The fastest $N=4$ model has the opposite problem, with a pattern speed high enough that the ILR and OLR are well inside the simulation radius (similar to bar simulations in the previous section) and there are no spiral arms in the outer disc.

The $N=2$ spirals with moderate pattern speeds (15-20\ps) show evidence of supplementary spiral structure branching off the main arms. The 15\ps\,model in particular has a pair of branches of comparable density to those driven by the spiral potential, but of a much shallower pitch angle (second panel, top, in Fig.\,\ref{CGArmPSxy}). These additional arm features are seen in other numerical studies of logarithmic spirals such as \citet{1994A&A...286...46P}, where the bifurcation of 2 to 4 armed spirals occurs at the inner 4:1 (ultraharmonic) resonance \citep{1997A&A...323..762P,2003ApJ...596..220C}. The additional branching arm features seem to peak in strength around 200\,Myr, and become less defined as evolution passes 500\,Myr.

The slowest $N=2$ models display very strong spur features inside of $R=7$\,kpc. For certain combinations of $N$ and $\Omega_{sp}$ (which determine the location of resonance features) we do not see gas tracing the spiral potential at low radii. Structure in the inner galaxy would need to be produced by the inclusion of a bar potential.

\subsubsection{Kinematic and radiative transfer \lv maps}
\begin{figure*}
\includegraphics[trim =30mm 10mm 0mm 0mm,width=180mm]{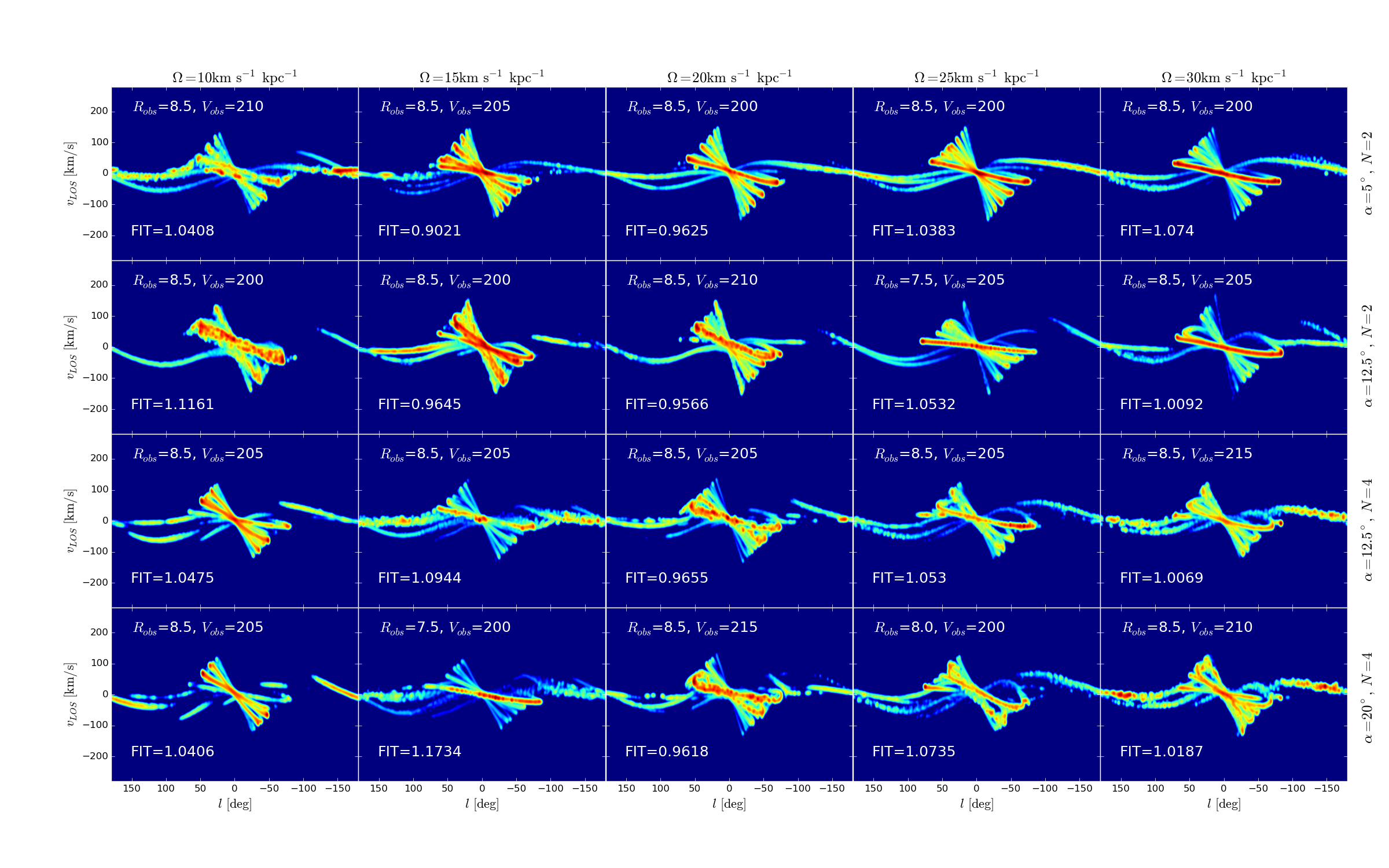}
\caption{The best fit \lv maps for the arm model of \citet{2002ApJS..142..261C} rotating at pattern speeds of 10, 15, 20, 25 and 30\,\ps\, increasing from left to right with pitch angle increasing from top to bottom (5\arcdeg, 12.5\arcdeg\,and 20\arcdeg). The values for the observer distance, circular velocity, and fit statistic are overplotted on each $\Omega_{sp}$-$\alpha$ pair (in kpc and \kms\,respectively). The maps are created after the simulation has evolved for 354\,Myr. The $\alpha=12.5^\circ$ models include both $N=2$ and $N=4$ morphologies. The 10\arcdeg\,and 15\arcdeg\,models are not shown but differ marginally compared to the 12.5\arcdeg\,maps.}
\label{CGArmLV}
\end{figure*}

A selection of \lv maps made using the the method described in section \ref{KinematicMapping} are shown in Fig.\,\ref{CGArmLV}. We show maps for $\alpha=$ 5\arcdeg , 12.5\arcdeg\,and 20\arcdeg\,and omit those for 10\arcdeg\,and 15\arcdeg\,due their  similarity to the 12.5\arcdeg\,models. The upper rows show $N=2$ models and the lower $N=4$ models. The maps are the results of the fit to $R_{obs}$, $V_{obs}$ and $l_{obs}$ similar to the previous section for the isolated bar models. Best-fitting parameters for the observer position and velocity are overplotted on each map. We include no bias towards certain values of $l_{\rm obs}$ as we did for fitting to the bar to constrain $\theta_b$.

We allow the \lv features to be fit by any part of the gas disc, rather than make assumptions about which \lv features should be fit by certain structures in \xy space. We experimented with masking out emission of local material when fitting the arm models, however the ability of some models to produce off-arm local material would be muted by this, and so we retain the fitting to the entire map.

General trends in the fitting are seen for all arm models. The strong local emission in the second quadrant is often fit by a major arm in the gas. The Local arm material appears significantly stronger than that of Perseus and Outer arms in the CO \lv data, giving the fit a preference to fitting to local material over the Outer arm, despite the physical size of the Outer arm being considerably greater. Fitting to the Local arm feature in \lv space causes the fit to miss the Outer arm in the second quadrant for $N=2$ models as there is simply not enough arm structure to produce 3 distinct arms in the first and second quadrants.

\begin{figure}
 \includegraphics[trim = 10mm 0mm 0mm 0mm,width=89mm]{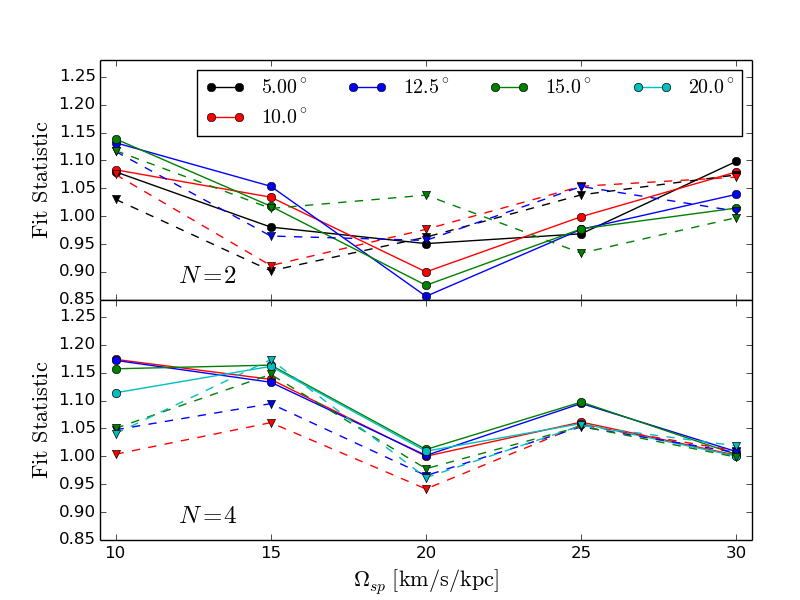}
  \caption{Fit statistic found by varying observer coordinates as a function of pattern speed of all \citet{2002ApJS..142..261C} type arm models, with various values for the pitch angle. Two different timestamps are shown as solid (236\,Myr) and dashed (354\,Myr) lines. $N=2$ and $N=4$ models are shown in the upper and lower panels respectively.}
  \label{ArmPS}
\end{figure}

The full results of our fitting to the observer's position using simple chemo-kinematic \lv maps are shown in Fig.\,\ref{ArmPS} as a function of arm pattern speed. The top panel shows the fit statistic for $N=2$ models with $\alpha=5^\circ,10^\circ,12.5^\circ$ and $15^\circ$ and the bottom panel the fit to $N=4$ models with $\alpha=10^\circ,12.5^\circ,15^\circ$ and $20^\circ$. Only the results for the 236 and 354\,Myr timestamps are shown for clarity. We also looked at the 472\,Myr timestamp and the trends with the fit were similar. Our overall interpretation is that the 20\ps\,models offer the best fit to the CO \lv data for both the $N=2$ and $N=4$ models. This is well within the observational bounds and is an often used value in other numerical investigations \citep{2011MSAIS..18..185G}. While $\Omega_b=$20\ps\, produces the lowest fit statistic for the $N=2$ arms it is not as consistent over time as the $N=4$ models. 

The $N=2$ arms favour a minimum of 15\ps\,for the later timestamp. Upon inspection of the individual \lv and \xy maps for this model (Fig.\,\ref{CGArmPSxy} and \ref{CGArmLV}), it is apparent that the supplementary arm branches mentioned previously are the cause of this minimum. The branches are approximately 90\arcdeg\,out-of-phase with the spiral potential and are much more apparent at 354\,Myr than 236\,Myr. These branches have a much shallower pitch angle than those being directly driven by the potential and decay before reaching the outer disc. This increase in arm features in the $N=2$, $\Omega=15$\ps\,models at later times allows for the reproduction of Perseus, Outer and Local arm features, but does not produce as strong emission in the third quadrant as that of the $N=4$ models (seen by comparing the $N=2$ and $N=4$, $\alpha=12.5$ models). This lowers the fit statistic compared to the $N=2$, $\Omega \ne 15$\ps\,models in the top panel of Fig.\,\ref{ArmPS} at the later time stamp.

\begin{figure}
 \includegraphics[trim = 10mm 0mm 0mm 10mm,width=89mm]{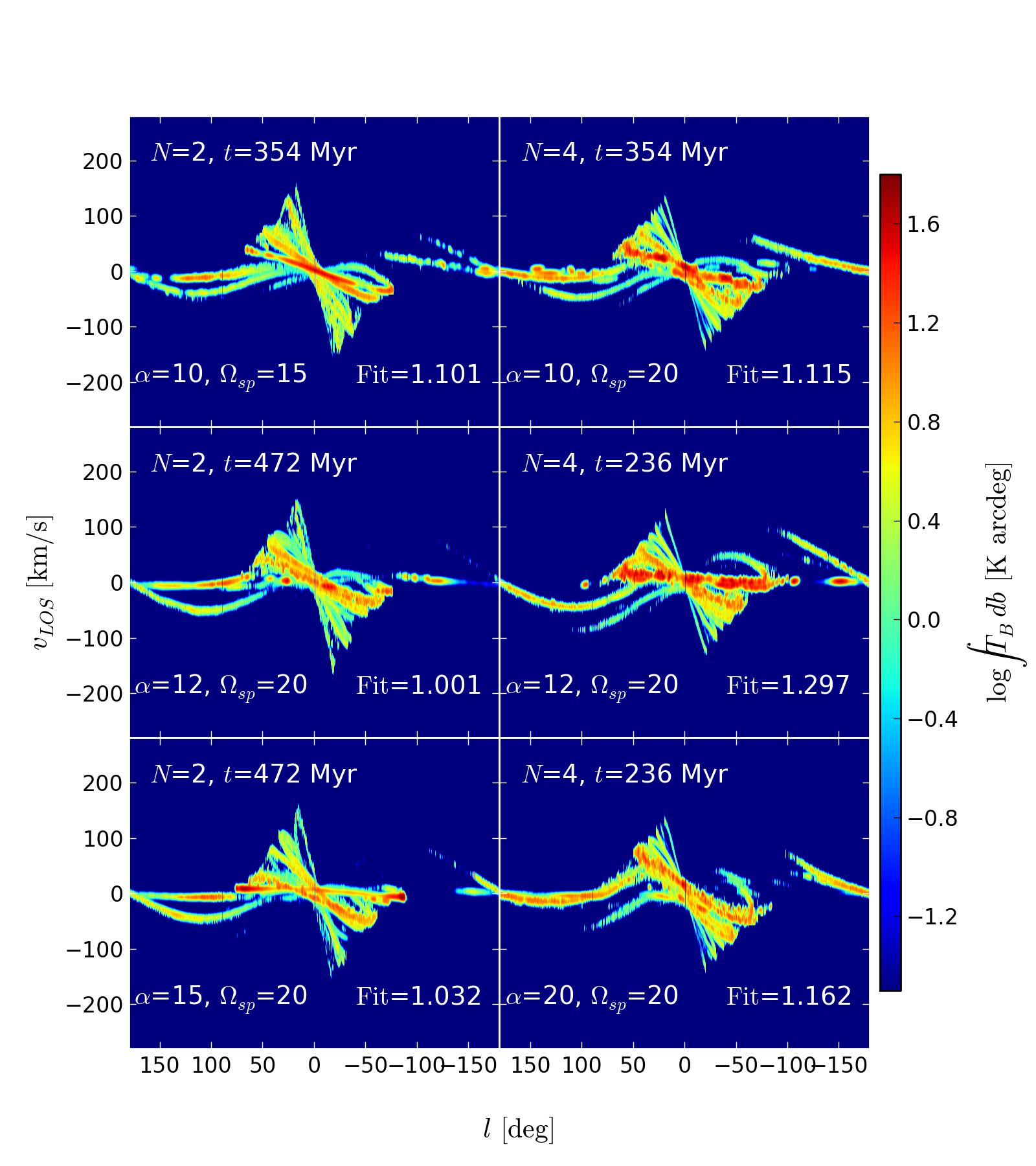}
  \caption{The best-fitting maps from isolated arm potentials for a variety of pitch angles. \lv maps are made using the \torus\, radiative transfer code, where the normalised fit to observed CO is shown in the bottom right of each panel. $\Omega_{sp}$ and $\alpha$ are in the units of \ps\,and degrees respectively shown in the bottom left.}
  \label{TorusArms}
\end{figure}

The number of spiral arms is perhaps the most important parameter driving the distribution of stars and gas in the Galactic disc. Fig.\,\ref{ArmPS} shows a slight preference towards $N=2$ over $N=4$ arm models. The $N=2$ models in the upper panel have a lower fit statistic minimum compared to the $N=4$ models, though there is also a greater spread in the former. Figure \ref{TorusArms} shows a selection of 6 of the best-fitting arm models made using \torus, each with a different combination of $N$, $\alpha$ and $\Omega_{sp}$. The $N=2$ models cover a reduced area of \lv space compared to their $N=4$ counterparts. This allows for $N=2$ models to match emission in the $2^{\rm nd}$ quadrant while leaving the  $3^{\rm rd}$ comparatively empty. This is seen in observations of CO, where possible arm features are much weaker in the in the $3^{\rm rd}$ quadrant compared to the $4^{\rm th}$ (Fig.\,\ref{DameLV}). The $N=2$ models tend to have the near arm aligned with the Perseus arm feature in the $2^{\rm nd}$ quadrant and this arm reaches the edge of the disc just as it enters the $3^{\rm rd}$ quadrant. The local emission in the $2^{\rm nd}$ quadrant is reproduced by interarm branches rather than the spiral arm that traces the potential, as it does in the best-fitting 4-arm models (as seen in the top panels of Fig.\,\ref{CGArmPSxy}).

The additional arm features in the $N=4$ models allow the reproduction of the 3 arm features seen in the observed CO data in the $1^{\rm st}$ and $2^{\rm nd}$ quadrants (Local, Perseus and Outer arms). They are also able to reproduce the characteristic ``hook" in \lv space from the Carina arm in the $4^{\rm th}$ quadrant while also placing material along the Perseus and Local arms. This is seen in the $N=4$, $\alpha=12.5$, $\Omega_{sp}=20$\ps\, model in Fig.\,\ref{TorusArms}. In order to fit to the Carina arm, there must be an arm structure placed very close to the observer's position. For pure logarithmic spirals with constant pitch angles this will result in very bright horizontal structures in \lv space, as seen in Fig.\,\ref{TorusArms}. This is clearly at odds with the observed emission in CO (and HI), which contains no strong emission at local velocities in the inner Galaxy. There was no single arm model that could place local emission, the Carina arm and the Perseus arm in their correct places, as well as producing a strong ridge of emission angled correctly in the inner Galaxy. From Fig.\,\ref{TorusArms} it can be seen that for any model that has a central ridge that is similar to that seen in CO observations, the Carina arm-like structure is pulled into the $|v_{los}|<20$\kms\,range. The resulting arm emission from the $N=4$ models in the $3^{\rm rd}$ quadrant is detrimental to the goodness of fit, due to the lack of molecular emission in the observations. This excess emission makes the $N=4$ models systematically worse compared those with $N=2$.

Out of all parameters the pitch angle of the arms is the poorest constrained in our arm-only models. Figure \ref{ArmPS} shows no strong preference towards any given pitch angle, in the 2-armed case especially. The minima of all arm models are at 12.5\arcdeg\,and 10\arcdeg, both of which have pattern speeds of 20\ps. At this stage there may simply be too many variables to establish a best-fitting pitch angle, especially when the orientation of the arms is still a completely free parameter (determined by the best-fitting $l_{\rm obs}$). The pitch angle produces fairly subtle differences in morphology compared to the arm number and pattern speed, which could explain the relatively loose correlations seen in Fig.\,\ref{ArmPS}. To try to find a stronger fit to $\alpha$ we attempted to fit to only the outer quadrants, where the arms should dominate the \lv structure, and negate the dominance of the central ridge in the fit statistic. The results were still inconclusive, and the fit behaved similarly as it did to the entire Galactic plane.

A full fit to the all features in \lv space seems impossible without the inclusion of a strong bar to drive additional features in the inner disc, allowing the arms to produce the Carina and Perseus features in the outer quadrant without trying to fit the central ridge simultaneously. The placement of the OLR of the bar at roughly the solar position would also impact upon the structures observed in the $1^{\rm st}$ and $4^{\rm th}$ quadrants. 

To further narrow down our parameter space for simulations with both arm and bar potentials we reject our $\alpha=$5\arcdeg\,and 20\arcdeg\,models. By-eye inspection shows that while these models do cover a similar area of \lv space as observations, they do not trace the features correctly. The 5\arcdeg\, models appear similar to concentric rings in \lv space, with many bright tangencies along the terminal velocity curve. The 20\arcdeg\,models appear too wide to match features in \lv space, and stray from the potential structure at $R>9$kpc. As there is no clear preference towards a 2 or 4 armed model seen for isolated arm simulations, we continue to use both 2 and 4-armed models in conjunction with the best bar models from the previous section. We choose to primarily use the minimum from Fig.\,\ref{ArmPS} of $\Omega_{sp}=20$\ps\,for further arm simulations. We also include 2-armed, $\Omega_{sp}=15$\ps potentials due to the secondary minimum in Fig.\,\ref{ArmPS}.

\subsubsection{Arm strength and model type}
\label{ArmModels}
\begin{figure}
 \includegraphics[trim = 15mm 0mm 0mm 0mm,width=93mm]{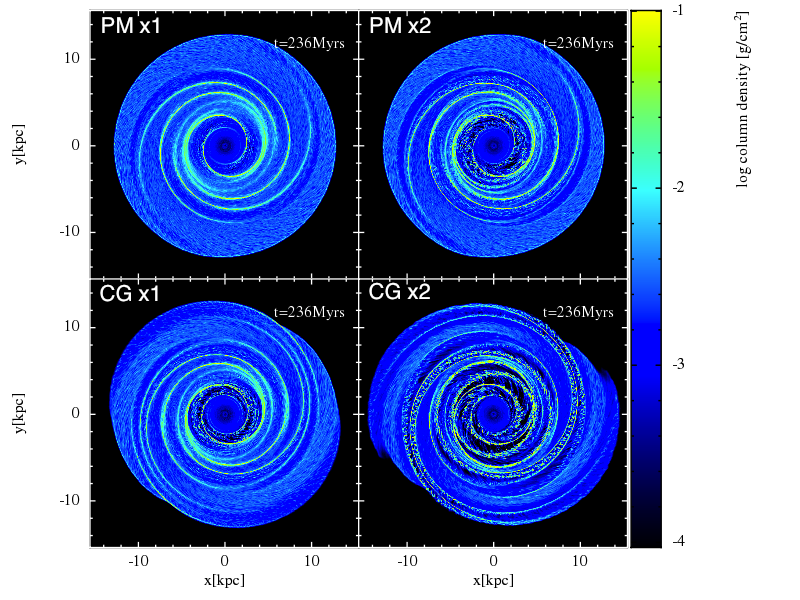}
  \caption{Multiple 2-armed models with different strengths. Top: the \citet{2003ApJ...582..230P} models, bottom: the \citet{2002ApJS..142..261C} models. The right hand panels have a strength $\times$2 the fiducial value used in this work. The potentials have a pattern speed of $\Omega_{sp}=20$\ps, pitch angles of $\alpha=12.5^\circ$ and are shown after 236\,Myr of evolution.}
  \label{PMbar}
\end{figure}

\begin{figure}
\includegraphics[trim = 10mm 0mm 0mm 0mm,width=90mm]{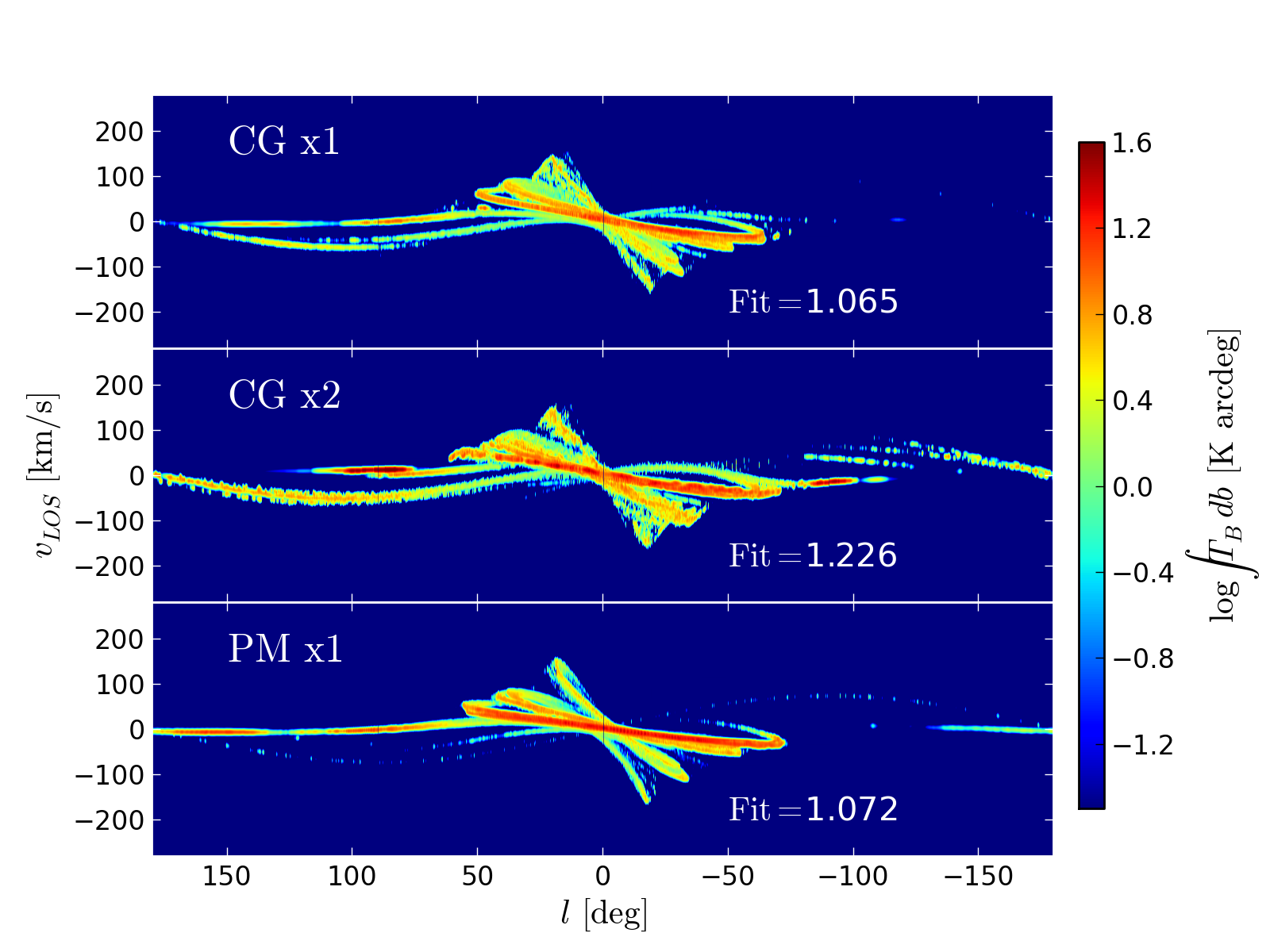}
\caption{Radiative transfer \lv maps constructed using \torus, of our different arm models; \citet{2002ApJS..142..261C} with normal and double strength, and \citet{2003ApJ...582..230P} arms. All models have the parameters $N=2$, $\alpha=12.5^\circ$, $\Omega_{sp}=20$\ps and have evolved for 236\,Myr. The corresponding top-down maps are shown in Fig.\,\ref{PMbar} where the observer is located at $y=8$\,kpc with a circular velocity of 210\kms (the best-fitting values for the CG$\times$1 arms).}
\label{PMmaps}
\end{figure}

In addition to the standard CG spiral arms we performed calculations with arm potentials with double the strength. Some of the models shown in Fig.\,\ref{CGArmPSxy} have arm features that do not appear to drive a high density spiral structure in the gas, especially the $N=2$ models. The bottom panels in Figure \ref{PMbar} show a comparison between a standard and double strength $N=2$ potential. Increasing the strength of the potential results in much clearer arm features. Characteristic 4:1 and 2:1 resonant orbits become clear much earlier in this simulation due to the increase in strength of the potential. In Figure \ref{PMmaps} we show the corresponding \lv maps created using \torus\,(x1 and x2 strength models in first and second panels respectively). The broad emission features appear effectively the same as their normal strength counterparts, only varying slightly in the interarm regions. This is expected as the production of CO is capped by the abundance of C given to the simulation. Raising the gas density in the spiral arms will not incur a much greater increase in emission, except by the accumulation of additional gas particles. Our fiducial arm strengths produce gaseous arms that are clearly visible in \lv emission maps (see Fig.\,\ref{TorusArms}) perhaps even too strong in  the $3^{\rm rd}$ quadrant. We conclude that there is no need for the higher strength models as they provide little advantage to the standard strength potentials.

We also ran simulations with the PM arm models with $\Omega_{sp}=20$\ps\,and $\alpha=12.5^\circ$, shown in the top panels of Fig.\,\ref {PMbar}. We used spiral masses of $1.5\times10^9\, \rm M_\odot$ (left-hand panel) and $2.6\times10^9\, \rm M_\odot$ (right-hand panel), where in \citet{2003ApJ...582..230P} the authors state the supplementary arm features are stronger in the lower mass case. Fig.\,\ref{PMbar} shows that the PM arms do indeed drive additional arm structures, appearing strongest in the mid-Galactic disc. These additional spiral branches have shallower pitch angles than the arms driving their formation, and are nearly circular approaching the solar radius. At later times the PM resonant arms become less pronounced, and the 4:1 resonance begins to dominate the flow of gas around $R=6$kpc (the same position as the ILR of 4-armed models). Arm branches are also present in the CG models (lower panels in Fig.\,\ref{PMbar}). The branches in the PM arms are slightly stronger than those seen in the CG potential, but the primary arms in the PM model are relatively weaker than those of the CG potential. The \torus\,map of the fiducial strength PM arms is shown in the bottom panel of Fig.\,\ref{PMmaps}. The branches appear clearly in \lv space, with emission of comparable strength to the arms. The lack of strong arm features in the PM models outside 9\,kpc makes it impossible for these arms to show the Outer and Perseus arm emission behind the observer in the $2^{\rm nd}$ quadrant. While the PM arm model is effective at creating 4-armed gaseous distributions from only a 2-armed potential, we do not find it suitable for re-creating all spiral features seen in \lv space. As a result, we do not perform any calculations with this potential including the effects of the bar.

\subsection{Simultaneous arm and bar simulations}
\label{MixModels}
Once a more refined parameter space had been selected we performed simulations with both bar and arm potentials simultaneously, using the CG, WK and WKr2 potential models. Parameters in bold in Table \ref{POTparams} are those used in arm-bar simulations, chosen based on fits in previous sections. Note that we use $\Omega_b=50$\ps\,for the WK and $\Omega_b=60$\ps\, for the WKr2 potentials. We use $\Omega_{sp}=15$\ps, $N=2$ arms only in conjunction with the $\Omega_b=60$\ps\,bar potential as the OLR of the $\Omega_{b}=50$\ps\,bar is close to region of arm branching and this may result in a disruption of these features.

Some of the resonances occupy the same radii in the above ranges. For example, a $N=4$ spiral at $\Omega_{sp}=
20$\ps\, and a bar with $\Omega_b=50$\ps\, has the ILR of the arms at approximately the same radius as the OLR of the bar, implying that a clear distinction between arm and bar features should be seen in this model. In general the bar CR will lie between the arm ILR and CR for $N=2$ models, but not for $N=4$ models.

\begin{figure*}
\includegraphics[trim =10mm 0mm 0mm 0mm,width=185mm]{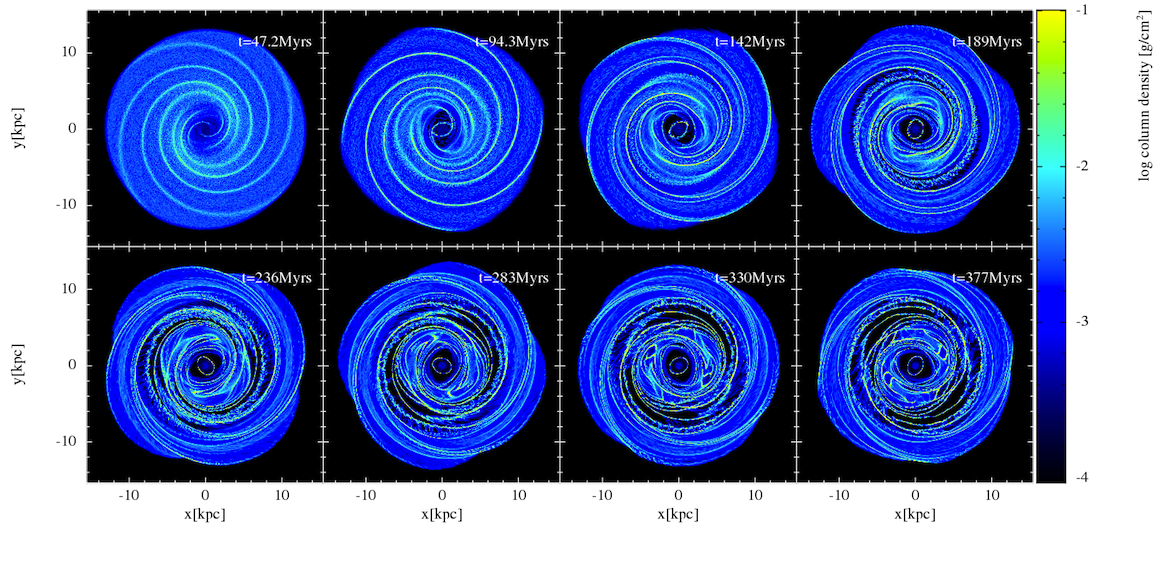}
\caption{Example of the evolution of a barred-spiral Milky Way simulation. The central bar is of WK type with and the arms of CG type. The potential parameters are; $N=4$, $\alpha=12.5^\circ$, $\Omega_{sp}=20$\ps\,and $\Omega_b=50$\ps.}
\label{Mix4Arm}
\end{figure*}

\begin{figure*}
\includegraphics[trim =10mm 130mm 0mm 0mm,width=185mm]{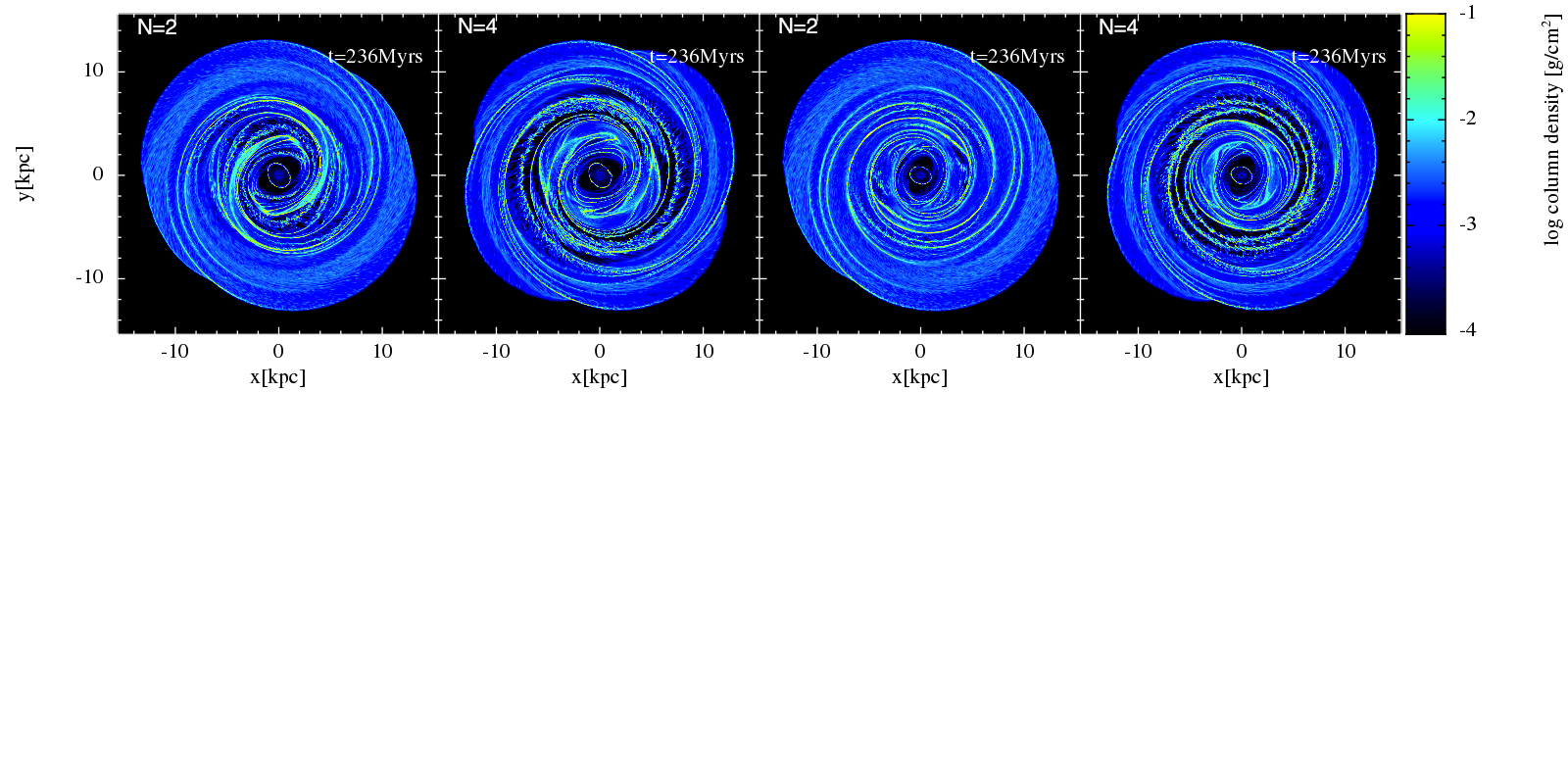}
\caption{Top-down maps of the gaseous response to the different $N-\Omega_b$ potential pairs, all of which have $\alpha=12^\circ$, $\Omega_{sp}=20$\ps\,and evolved for 236\,Myr. The bar potential in the left-hand panels has $\Omega_{b}=50$\ps, and $\Omega_{b}=60$\ps\,in the right-hand panels, shown in conjunction with 2 and 4 armed models.}
\label{MixPatternSpeed}
\end{figure*}

\subsubsection{Simulation \xy maps}
An example of the evolution of a barred-spiral simulation is shown in Fig.\,\ref{Mix4Arm}, with the parameters; $N=4$, $\alpha=12.5^\circ$, $\Omega_{sp}=20$\ps\,and $\Omega_b=50$\ps\,(with CG and WK type potentials). The addition of a bar distorts the arm features within a radius of 5\,kpc, roughly corresponding with the bar's OLR. The bar-arm contact region has a large amount of complex structure where the gas in the arm potential strays from a logarithmic spiral structure to join those arms driven by the bar which are much tighter wound. After 500\,Myr the gas around the bar establishes elliptical orbits similar to those seen in Fig.\,\ref{WDBarPSxy}, though the addition of arm potentials inhibits the formation of parallel and perpendicular elliptical orbits seen at the OLR in bar-only simulations. We find that, as suggested by \citet{1988MNRAS.231P..25S}, there is a clear inner region dominated by the bar potential and outer region dominated by the spiral potential, with only a small region where the two are intermixed.

The differences between the models as a function of $\Omega_b$ and $N$ are shown in Fig.\,\ref{MixPatternSpeed}. The slower bars disrupt the arm features up to the solar radius, while the faster bars are less radially extended, allowing arms to approach smaller radii. The 2 armed models still have a dearth of high density interarm material, though the arms in conjunction with the slower bar has additional interarm structure caused by the large radial extent of the features driven by the bar (though this is more evident at later times).

An additional complication to the barred-spiral models is the offset between the arm and bar potentials, which is time-dependent due to $\Omega_{sp} \ne \Omega_b$. By choosing to analyse the model at specific timestamp, as in the arm and bar only simulations, we would have already selected the offset between the bar and arm features. Instead, we analysed each barred-spiral model in the range of 280-370\,Myr, regardless of arm number and bar pattern speed. This range was the minimum required time between arm passages around a reference frame aligned with the bar for all models considered and includes the full possible range of arm-bar offsets. The main difference over this time-frame is the location and amount of interarm material. 

\subsubsection{Kinematic and radiative transfer \lv maps}
\begin{figure*}
 \includegraphics[trim = 10mm 10mm 0mm 5mm,width=180mm]{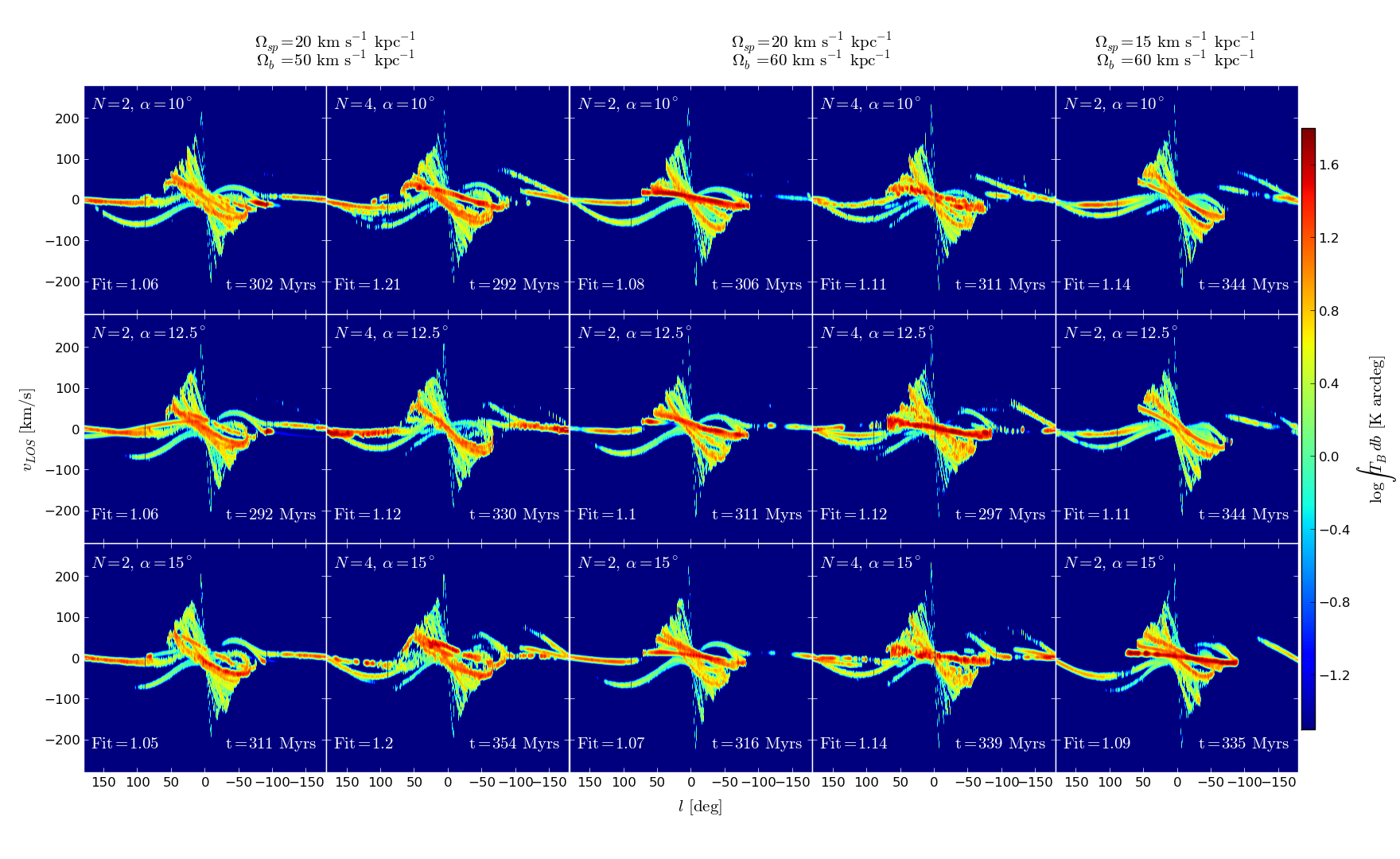}
  \caption{Synthetic emission maps made using \torus\, for our barred-spiral models with $\theta_b=45^\circ$. The arm position relative to the bar is found using the method of fitting to the observer coordinates in the isolated arm and bar cases. The first two columns show $\Omega_b=50$\ps\, with $N=2,4$ respectively, and the second two show $\Omega_b=60$\ps\, with $N=2,4$. The fifth column has a slower arm pattern speed of  $\Omega_{sp}=15$\ps. The spiral arm pitch angle increases from top to bottom.}
  \label{MixTorus}
\end{figure*}

The \torus\, emission maps for each $N$-$\Omega_b$-$\alpha$ combination are shown in Figure \ref{MixTorus} with the best-fitting values of $R_{\rm obs}$, $V_{\rm obs}$ and arm-bar offset (i.e. evolution time) found using the method described in Section \ref{KinematicMapping}. We have fixed the bar at $\theta_b=45^\circ$, which is consistent with the best-fitting value found in our bar-only simulations, to allow a reference point for altering the arm-bar offset. Simple by-eye comparisons between these maps shows that whilst most fit some features well ultimately none shown a perfect match to the data, suffering the same problems as the arm-only models in Section \ref{ArmSec}. As was the case in the arm only models, the fit statistic is uncorrelated with the pitch angle. If the fit statistic is averaged across all parameters except pitch angle then there is a marginal preference towards $\alpha=12.5^\circ$. There is also a preference towards a pattern sped of $\Omega_b=50$\ps\, for $N=2$ models and $\Omega_b=60$\ps\, when $N=4$.

\begin{figure*}
 \includegraphics[trim = 0mm 0mm 20mm 0mm,width=160mm]{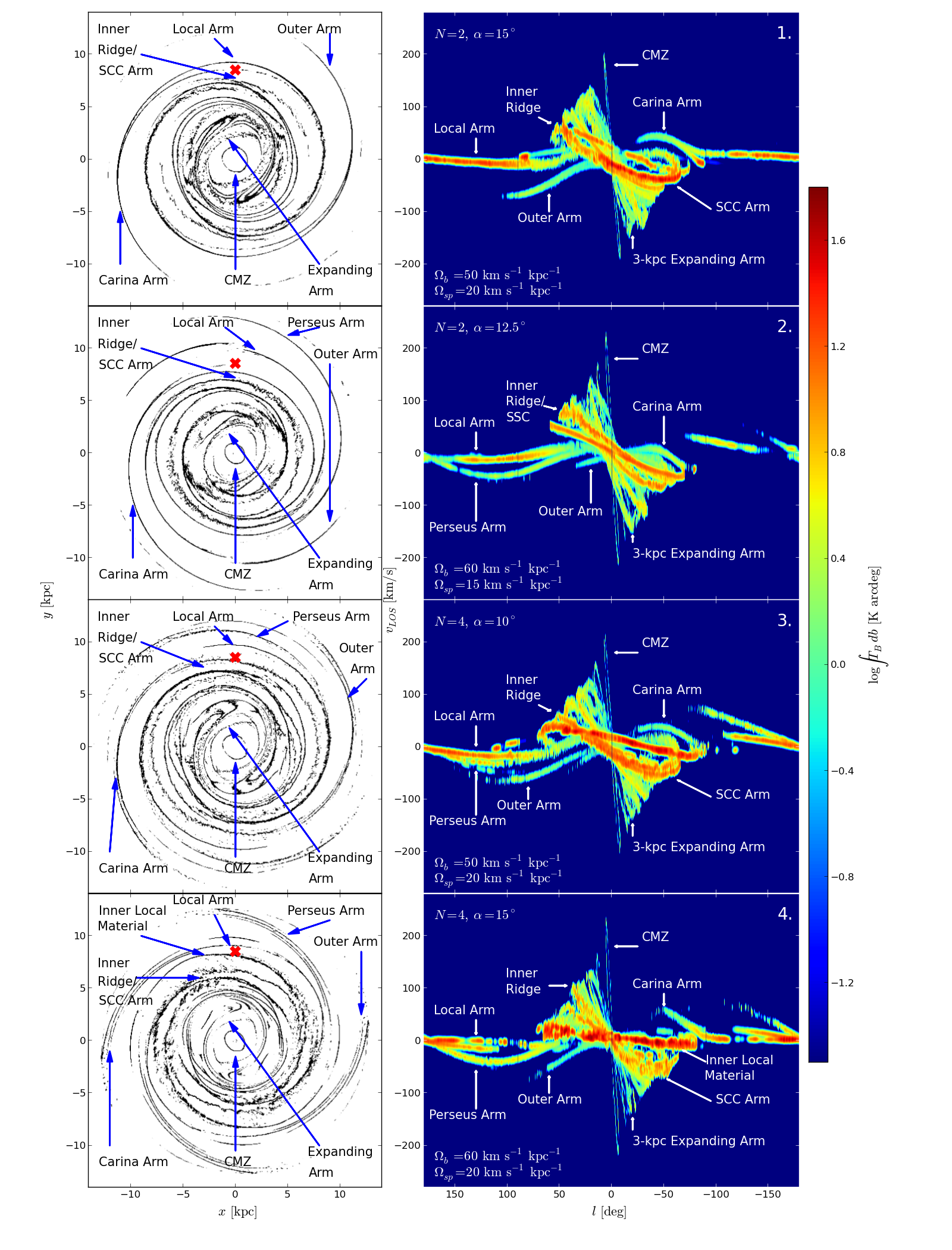}
  \caption{Four CO radiative transfer \lv maps with their \xy counterparts from Fig.\,\ref{MixTorus}, chosen to show a range of different morphologies. The top-down maps  only show material that is seen in CO \lv space; that of the highest density. The cross indicates the observers position (which differs between models). SCC refers to the Scutum-Centaurus-Crux arm in the 4-armed paradigm of the Milky Way, also referred to in the main text as the Inner Ridge. Arrows indicate locations of prominent features in \lv space. Models 2 and 4 reproduce the outer arm structure while 1 and 3 provide a better reproduction of the Carina arm.}
  \label{MapLabel}
\end{figure*}

The reasons preventing a good fit to all emission features are covered by the following examples.
In Fig.\,\ref{MapLabel} we show four different arm-bar simulations from Fig.\,\ref{MixTorus} in both \lv and \xy space. These have been chosen to highlight the main differences between the simulations, and are not necessarily the best fits from Fig.\,\ref{MixTorus}. In the first panel we show a 2-armed spiral model with our slower bar (50\ps). The \lv map in this case shows a good reproduction of the Carina arm, and Local arm material in the second quadrant (this is common to all 2-armed model fits in Fig.\,\ref{MixTorus}). The \xy map shows that the \lv Carina arm feature in this model actually joins with the Local arm material. The Carina segment branches away as it nears the solar position, passing though $R<R_o$ while the Local arm feature breaks away from the spiral potential and maintains a radial distance of $R>R_o$ upon passage into the first quadrant. The major drawback of this and other 2-armed models is the failure to produce the Outer, Perseus and Local arms simultaneously. Two armed-models produce an inner emission ridge seen in observations (a combination of the Scutum-Centaurus-Crux, SCC, arms and possibly a molecular ring). However, the ridge in this case is too shallow in \lv space, implying it is too close to the Solar position.

In the second panel, we show another 2-armed model with a moderate pitch angle (12.5\arcdeg) and a slow bar (50\ps), but with a slower arm pattern speed than the previous model (15\ps). This value of $\Omega_{sp}$ provides strong branching features that can be seen in the \xy map, driving a 4-armed gas structure from only a 2-armed potential. This model reproduces the Perseus, Outer and the Local arms. Reproducing these arm features simultaneously would be impossible for normal a 2-armed structure (as in the previous model). The Local and Outer arms are actually reproduced by the branches, not the arms directly tracing the potential. The SCC arm/inner ridge is angled similarly to Fig. \ref{DameLV}, and the 3kpc-expanding arm is very clearly seen in \lv space. The main flaw in this model is the position of the Carina arm, which does not reach into the $v_{los}>0$\kms region as seen in observations.

The third panel shows a 4-armed model with a shallow pitch angle (10\arcdeg). In this case, there is clear reproduction of the Carina arm feature, located inside the Solar radius in \xy space. As this feature passes in between the Solar position and the Galactic centre it causes a bright emission feature at near-local velocities, a feature not seen in observations. The SCC arm feature is seen behind this strong emission feature in \lv space. The second quadrant arm features are not as clear as the previous model, with the Local and Perseus features not clearly separated in \lv space. The feature here labelled as the Outer arm could equally be labelled the Perseus arm, but would leave multiple arm structures unidentified in the outer Galaxy, caused by a large amount of branching material in the $7{\rm kpc}<R<11{\rm kpc}$ region seen in \xy space.

The final panel also shows a 4-armed model, with a wide pitch angle (15\arcdeg), but with a faster bar than the previous panels (60\ps). The faster bar is less extended radially, allowing the gas to trace the spiral potential to smaller radii. In the \xy map the spiral arm pitch angle is maintained to $R\approx 4$kpc, whereas in the slower, 50\ps, models in the upper panels structure is dominated by the bar until $R\approx 6$kpc. This model appears to produce all the observed features; Local arm, Perseus arm, Outer arm, SCC arms/ridge and Carina arm. The problem again is that arms must pass in front of the observer to appear in the fourth quadrant, producing emission that dominates the SCC feature in the inner Galaxy. This model in particular has little emission in the third quadrant, as seen in observations, owing to the Perseus arm disappearing as it leaves the second quadrant. The Carina arm feature is located at higher values of $v_{los}$ than is seen in observations, however there are similar maps for the $\alpha=12.5^\circ$ case that provide a better match for this section, but are not shown in this figure.

\section{Discussion}
\label{Discussion}
The models shown in figures \ref{MixTorus} and \ref{MapLabel} show it is possible to reproduce all features of the \lv data. However, we find it difficult to produce a good match to all features simultaneously.

\begin{figure}
 \includegraphics[trim = 20mm 0mm 60mm 10mm,width=70mm]{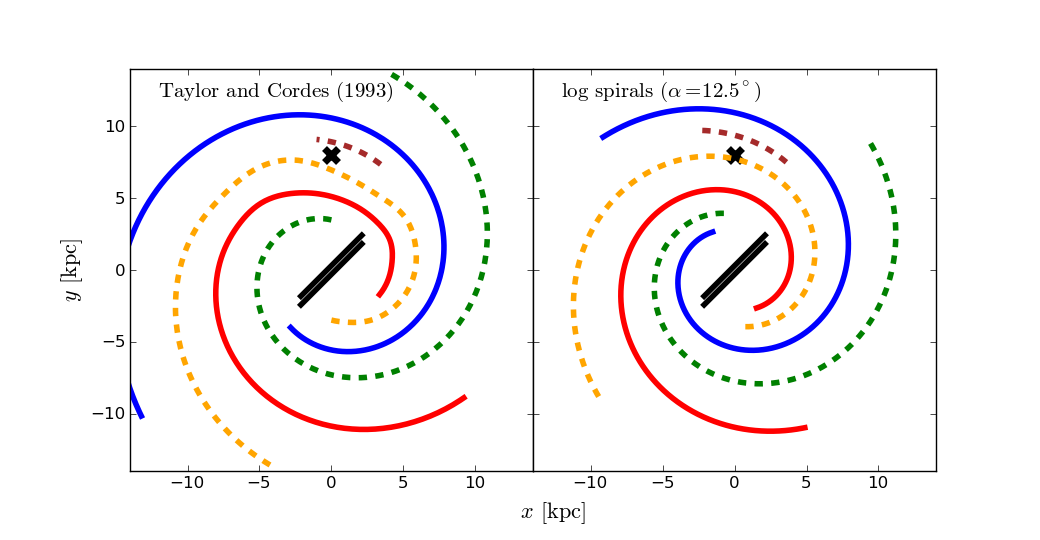}
 \includegraphics[trim = 20mm 10mm 60mm 5mm,width=70mm]{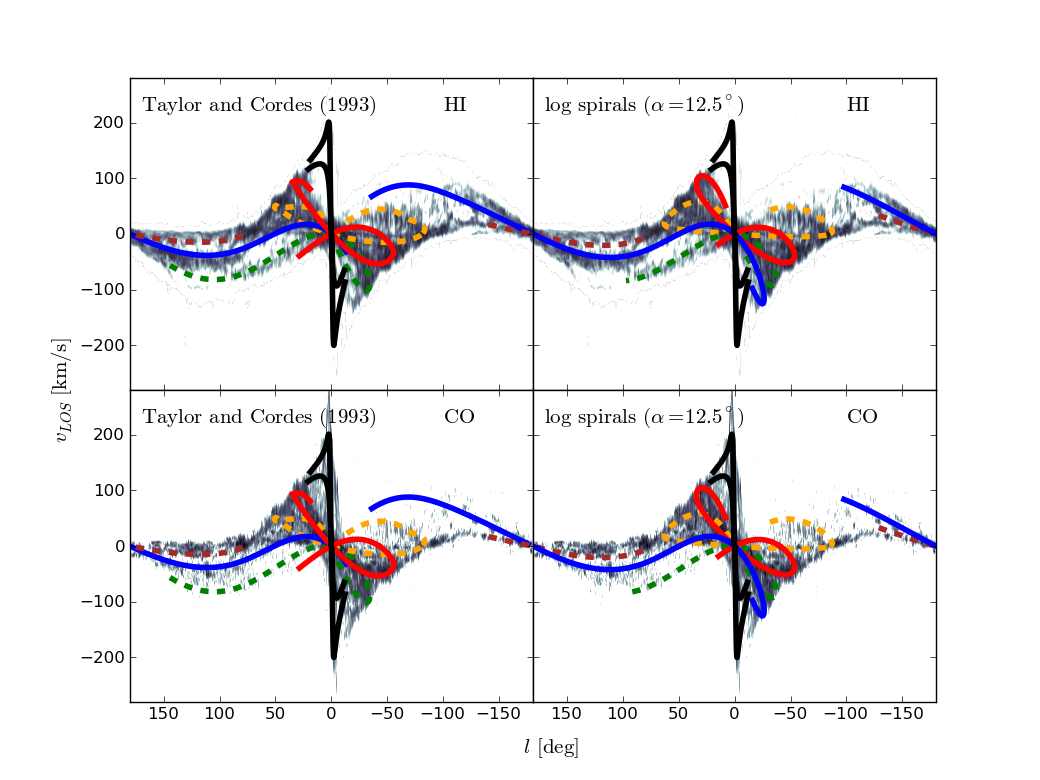}
  \caption{Different arm models in \xy plane (top figure) and their projection onto \lv space (bottom figure). Left panels: arm model of modified logarithmic spirals from \citet{1993ApJ...411..674T}, primarily constrained to data from \citet{1976A&A....49...57G}. Right panels: simple 4-armed spiral model with each arm offset by $\pi/2$ from the previous with addition of a local arm segment. Arms only extend radially to distance required to match \lv emission features. Observed CO and HI emission data is plotted on grey-scale behind the model arm features in the lower figure. Bold lines indicate the strong primary arm features in the old stellar population inferred by \citet{2009PASP..121..213C}.}
  \label{Models}
\end{figure}

Four armed models are more capable of fitting multiple features simultaneously, but to do so must place some arm structure just inside the solar position. This must be within very close proximity to allow the tangent point of the Carina arm to reach out to $l\approx -90^\circ$. While a strong emission feature is seen in the inner Galaxy in observations, it is angled much steeper in \lv space than our synthetic maps. One can conclude that the local SCC arm material is either lacking in molecular material or that the shape is far from that of a logarithmic spiral near the Solar position. If it is indeed lacking in molecular gas, then it can be expected to at least be rich in atomic gas. The HI \lv observations show much more structure at local velocities in the inner Galaxy, which could be the SCC arm features that are not seen in CO (the HI \lv map is shown in Fig.\,\ref{Models}). 

Alternatively, the Carina-Sagittarius arm structure could deviate significantly from a normal logarithmic-spiral structure. This is supported by other works in the literature (e.g. \citealt{1976A&A....49...57G}, \citealt{2008A&A...486..191P}). These models involve some straight section of the SCC arm as it passes in front of the observer. In Fig.\,\ref{Models} we show such a model, specifically that of \citet{1993ApJ...411..674T}, compared to a 4-fold symmetric spiral pattern similar to that used in this study. This additional distance between the observer would give the arm a greater line-of-sight velocity, pulling it up and away from the $V_{\rm obs}=0$\kms\, line in our maps in Fig.\,\ref{MixTorus}, as seen in projection in the bottom left panel of Fig.\,\ref{Models}. It is also seen in observations that while the Sagittarius and Carina tangents are well traced by distance determinations, there is very little material placed on these arms in the local Galaxy in the direction of the Galactic centre (e.g.  \citealt{1976A&A....49...57G}, \citealt{2003ApJ...587..701F}, \citealt{2003A&A...397..133R}, \citealt{2009A&A...499..473H}). It also may be that the arm structure is better represented by a transient and irregular spiral structure, rather than that of a fixed grand design galaxy. These structures are reproducible in simulations through the inclusion of a live stellar disc, rather than fixed analytical potential (e.g. \citealt{2009ApJ...706..471B,2010MNRAS.403..625D,2012MNRAS.426..167G}).

In all of our \lv maps we fail to reproduce the structure of the CMZ. In certain instances we do produce velocities that are comparable to the highest values seen in observations, for example those in the upper panels of Fig.\,\ref{WDBarPatternLV}. The peak velocity structures in our models stem from the inner $x_2$ bar orbits perpendicular to the bar major axis, and appear as a symmetric loop structure in \lv space, while the observed CMZ is highly asymmetric. The SPH particles that are present have aligned themselves with the $x_2$ orbits, leaving little material available to fill in the missing emission. In order to fully capture the asymmetric emission features in the central galaxy a dedicated simulation is required of only the inner galaxy to better resolve the gaseous features. The addition of stellar feedback or a live stellar disc may also be required to break up the symmetric inner bar orbits.

In \citet{2004ApJ...615..758G} the authors construct synthetic \lv maps by simply mapping structures in \xy on-to \lv coordinates. They too show that while the Carina ``hook" is easy to reproduce, it causes a strong dense ridge angled far too shallow in \lv space compared to that seen in observations. They also note that crowding in velocity space can cause ridges in \lv space not necessarily corresponding to high density gas regions. As CO traces high density regions only we do not have that problem here, and our \lv features correspond well with high density gas regions associated with arm and bar features.
Our results are at some odds with the work by \citet{refId0}, who find that a bar pattern speed of 30\ps\,is the best match to the \lv diagram, without the inclusion of arm potentials. Our value is more in keeping with that suggested by \citet{1999A&A...345..787F} and \citet{1999MNRAS.304..512E}. Our lower pattern speed of 40\ps\,resulted in extremely strong emission in front of the observer, features that would not appear in the aforementioned works due to the mapping of \xy features to \lv space lacking a radiative transfer treatment.

There are further observational constraints that we do not include here. These include measuring the rotation curve, comparing with \lv maps of HI (e.g. \citealt{2005A&A...440..775K}) and the bar driven velocity field (e.g. \citealt{1998MNRAS.298..387D}).

\section{Conclusions}
\label{Conclusion}
We have used smoothed particle hydrodynamics and radiative transfer codes to create the first synthetic emission maps of the CO emission of our Galaxy. We obtain good agreement of the values of CO emission in our synthetic maps compared with those of \citet{2001ApJ...547..792D} but find that CO emission is quite sensitive to the mean density of the gas (see also Duarte-Cabral et al. in preparation).

We then use the CO maps created from a large number of simulations to try and determine the spiral arm and bar morphology of the Milky Way. By comparing maps of simulated CO emission in \lv space with the Dame CO map, we identify which parameters (including pattern speeds, number of spiral arms, pitch angles, a number of different arm and bar potentials, and the position of the observer in the Galaxy) produce the best-fitting synthetic CO map. Whilst other authors have produced individual \lv maps from simulations, here we extend this idea to using multiple simulations to carry out a systematic study of the available parameter space. We perform a large number of calculations with bar, spiral and both bar and spiral potentials, assuming that the Galaxy is of grand design with logarithmic spiral arms of constant pitch angle and pattern speed. Our calculations are by design simple, in that although we include heating and cooling and basic ISM chemistry, we neglect gas self gravity and stellar feedback (which are computationally very expensive), in order to search a wide parameter space. 

For our simulations with just bars, although the bars do drive spiral arms, they fail to produce spiral structure in the Outer Galaxy. Likewise our models with only spiral arms fail to produce enough structure in the inner Galaxy. Some parameters gave relative clear best fit values (e.g. $\Omega_{sp}$) whilst others, such as the spiral arm pitch angle, were less well constrained. Overall our best-fitting models favour a bar pattern speed within 50-60\ps\,and an arm pattern speed of approximately 20\ps, with a bar orientation of approximately 45\arcdeg\,and arm pitch angle between 10\arcdeg-15\arcdeg. We also left the position of the observer as a free parameter, and found that our fits give good agreement with observed values (i.e. $R_{obs}=8.5$\,kpc, $V_{obs}=220$\,\kms). 
Our models were unfortunately not able to readily discriminate between models with 2 and 4 spiral arms, though we find it difficult to reproduce all the observed \lv features simultaneously with only 2 arms. 
We tested the hypothesis that the Galaxy may contain 2 stellar spiral arms, which drive a 4 arm pattern in the gas, including trying the potential suggested by \citet{2003ApJ...582..230P}. Using this potential we were unable to reproduce all the observed \lv features, because the extra resonance features (branches) were too weak at large radii to produce significant CO emission. The only 2-armed potentials that produced branches of sufficient strength to produce significant \lv structure were those with a specific pattern speed of 15\ps.

Our calculations included models with a combined bar and spiral potential, but even with these we could not satisfactorily reproduce the observed CO features. Whilst it was possible to reproduce features in emission that are seen in observations, such as the Perseus arm, Carina arm, inner ridge emission, local material and the outer arm, these features are not all reproducible simultaneously. The 2-armed models cannot reproduce all these features, yet the 4-armed models create too much emission locally. Assuming logarithmic spiral arms, in order to successfully match the Carina arm feature, an extremely strong emission feature must be placed near $v_{los}$ = 0\kms\,in the inner Galaxy. Models which do not use radiative transfer may miss the significance of this feature. Alternatively the Carina arm would need to exhibit an irregular shape in the vicinity of the Sun. This leads us to the conclusion that while the 4-armed symmetrical model can produce many of the features seen in the \lv observations, it may be necessary to allow an irregular arm structure to convincingly match the Galaxy.

An alternative approach to that in this paper is to model the Milky Way as a transient, multi-armed galaxy by the inclusion of a live stellar disc. A study of the Milky Way ISM \lv emission using a live-stellar disc, and the comparison to the grand design case, will be the subject of a future study.

\section*{Acknowledgements}
We thank an anonymous referee, whose comments and suggestions improved the paper. We also thank Tom Dame for providing access to the CO longitude-velocity data. The calculations for this paper were performed on the DiRAC Complexity machine, jointly funded by STFC and the Large Facilities Capital Fund of BIS, and the University of Exeter Supercomputer, a DiRAC Facility jointly funded by STFC, the Large Facilities Capital Fund of BIS and the University of Exeter. ARP is supported by an STFC-funded post-graduate studentship. CLD acknowledges funding from the European Research Council for the FP7 ERC starting grant project LOCALSTAR. DJP is supported by a Future Fellowship funded by the Australian Research Council (FT130100034). Figures showing SPH particle density were rendered using \textsc{splash} \citep{2007PASA...24..159P}.

\bibliographystyle{mn2e}
\bibliography{Pettitt_morphology1.bbl}

\clearpage

\appendix

\section[]{Axisymmetric Potentials}
\label{AppAxi}

Our three-component axisymmetric galactic potential is composed of a separate disc, bulge and halo based on that of \citet{2003ApJ...582..230P} and \citet{1991RMxAA..22..255A}. The disc component is the standard Miyamoto-Nagai form \citep{1975PASJ...27..533M} with a potential of
\begin{equation}
\Phi_d(\varpi,z) = \frac{GM_d}{(\varpi^2+[a_d+(z^2+b_d^2)^{1/2}]^2)^{1/2}},
\end{equation}
where $a_d$ controls the radial scaling and $b_d$ the vertical, and $\varpi^2=x^2+y^2$. The bulge is described by a spherical Plummer potential \citep{1911MNRAS..71..460P},
\begin{equation}
\Phi_b (r) =- \frac{GM_b}{\sqrt{r^2+r_b^2}},
\end{equation}
with $r_b$ controlling the radial scaling, and $r^2=x^2+y^2+z^2$. The spherical dark matter halo is taken from \citet{1991RMxAA..22..255A},
\begin{equation}
\begin{aligned}
 & \Phi_h (r) =   -\frac{GM_h(r)}{r} \\ 
 &-\frac{GM_{h,0}}{\gamma r_h}\left[   -\frac{\gamma}{1+(r/r_h)^\gamma} + \ln \left( 1+\left(\frac{r}{r_h}\right)^\gamma\right)          \right]^{r_{h,max}}_r,
\end{aligned}
\end{equation}
where $r_{h,max}=100$ kpc is the halo truncation distance and $\gamma=1.02$. The mass inside the radius $r$ of the halo is given by
\begin{equation}
M_h(r)=\frac{M_{h,0}(r/r_h)^{\gamma+1}}{1+(r/r_h)^{\gamma}}.
\end{equation}

\begin{table}
 \caption{Fixed galactic axisymmetric potential parameters used to reproduce the observed rotation curve.}
 \begin{tabular}{@{}lcc}
  \hline
  Term & Description & Value \\
  \hline
  $M_d$ & Disc mass   &$ 8.56 \times 10^{10} \, \rm M_\odot$ \\
  $M_b$ & Bulge mass &$ 1.40 \times 10^{10} \, \rm M_\odot$\\
  $M_{h,0}$ & Halo mass  &$10.7 \times 10^{10}\, \rm M_\odot$ \\
  $a_d$ & Disc radial scale length   &$5.30\, \rm kpc$ \\
  $b_d$ & Disc vertical scale length&$0.25\, \rm kpc$ \\
  $r_b$ & Bulge radial scale length &$0.39\, \rm kpc$ \\
  $r_h$ & Halo radial scale length   &$12.0 \, \rm kpc$ \\
  \hline
 \end{tabular}
 \label{RCparams}
\end{table}

The total axisymmetric potential is then simply given by $\Phi_{bhd}(\vec{r})=\Phi_b(r)+\Phi_h(r)+\Phi_d(\varpi,z)$, and accelerations are then calculated using the gradient of the potentials; $\vec{f}_{ext}=-\vec{\nabla} \Phi(\vec{r})$. The various axisymmetric potential parameters are fixed throughout all simulations to best match the rotation curve of the Milky Way (Fig.\,\ref{MWRC}) and are given in Table \ref{RCparams}. There are numerous other potential sets in the literature we could have chosen to represent  the axisymmetric component.

\newpage
\section[]{Resolution study}
\label{ResStudy}
\begin{figure}
 \includegraphics[trim = 15mm 0mm 0mm 0mm,width=89mm]{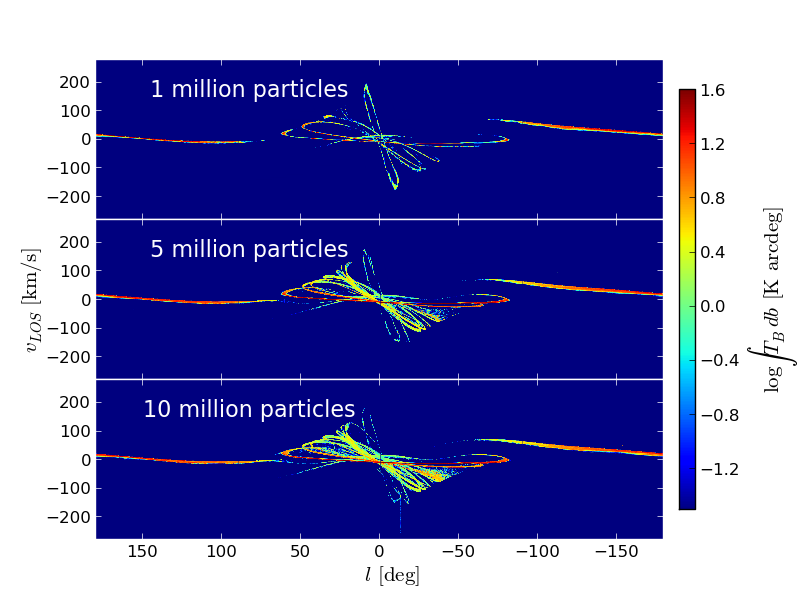}
  \caption{Radiative transfer CO \lv maps resulting from an SPH simulation with 1, 5 and 10 million particles (increasing from top). The gas is subject to a bar potential moving at 40\ps, shown after 280\,Myr of evolution. The observer is set to the IAU standard position and velocity. Turbulent velocity broadening is excluded to highlight differences between different resolutions.}
  \label{ResTest}
\end{figure}

To test our adopted simulation resolution of 5 million particles we run a number of simulations with 1 and 10 million particles. Top down maps of 1 million particles displayed significantly less structure around the resonance regions of the potentials, while 5 and 10 million calculations showed little difference. Figure \ref{ResTest} shows CO \lv emission maps made using \torus\,for simulations using 1, 5 and 10 million particles (increasing from top) inside an isolated bar potential. No turbulent velocity term is added to the line profiles so as to highlight the resolution effects. The difference between 5 and 10 million particles appears to be minimal, but the 1 million run has considerably less emission in the inner Galaxy in comparison. We conclude the 5 million particle resolution is sufficient to capture the global Galactic CO emission.

\bsp
\label{lastpage}

\end{document}